\documentclass{aa}  

\usepackage{graphicx}
\usepackage{txfonts}
\usepackage{color}
\usepackage{amsmath}
\usepackage{lscape}
\usepackage{pdflscape}
\usepackage{rotating}
\usepackage{subfiles}
\usepackage{stfloats}
\usepackage{graphicx}
\usepackage{txfonts}
\usepackage{color}
\usepackage{booktabs}
\usepackage{amsmath}
\let\oldAA\AA
\renewcommand{\AA}{\text{\normalfont\oldAA}}
\usepackage{amsfonts}
\usepackage{amssymb}
\usepackage{siunitx}
\usepackage{hyperref}
\usepackage{graphicx}
\usepackage{caption}
\usepackage{tabularx}
\usepackage{longtable, ltcaption}
\usepackage{pdfpages}
\usepackage{rotating, booktabs}
\usepackage{listings}
\usepackage{subfig}
\usepackage{fancyhdr}
\usepackage{lscape}
\usepackage{epsf}
\usepackage{changepage}
\usepackage[Sonny]{fncychap}
\usepackage{multirow}
\usepackage{scrextend}
\usepackage{alphalph}

\begin{document}

   \title{The WISSH quasars Project: \\ II. Giant star nurseries in hyper-luminous quasars}


   \author{F. Duras\thanks{E-mail: \texttt{federica.duras@oa-roma.inaf.it (OAR)}}
          \inst{1,2}
          \and
          A. Bongiorno \inst{2}
          \and
           E. Piconcelli \inst{2}
          \and
          S. Bianchi \inst{3}
          \and
          C. Pappalardo \inst{4,5}
          \and 
          R. Valiante \inst {2}
          \and
          M. Bischetti \inst {2}
          \and
          C. Feruglio \inst {6}
          \and 
          S. Martocchia \inst{7}
          \and
          R. Schneider \inst {8}
          \and 
          G. Vietri \inst {2}
          \and 
          C. Vignali \inst {9,10}
          \and 
          L. Zappacosta \inst{2}
          \and 
          F. La Franca \inst {1}
          \and 
          F. Fiore \inst {2} }

   \institute{Dipartimento di Matematica e Fisica, Università Roma Tre, via della Vasca Navale 84, I-00146, Roma, Italy 
         \and
             INAF Osservatorio Astronomico di Roma, via Frascati 33, 00040 Monteporzio Catone, Italy
           \and
           INAF Osservatorio Astrofisico di Arcetri, Largo E. Fermi, 5, 50125 Firenze, Italy
           \and
            Centro de Astronomia e Astrofísica da Universidade de Lisboa, Observatório Astronómico de Lisboa, Tapada da Ajuda
            \and
            Instituto de Astrofísica e Ciencias do Espaço, Universidade de Lisboa, OAL, Tapada da Ajuda, 1349-018 Lisboa, Portugal
            \and
            INAF - Osservatorio Astronomico di Trieste, via G. Tiepolo 11, I-34124 Trieste, Italy
            \and
            Astrophysics Research Institute, Liverpool John Moores University, 146 Brownlow Hill, Liverpool L3 5RF, UK
            \and
            Dipartimento di Fisica Università di Roma La Sapienza, I-00185 Roma, Italy
            \and
            Dipartimento di Fisica e Astronomia, Università di Bologna, viale Berti Pichat 6/2, I-40127 Bologna, Italy
            \and
            INAF - Osservatorio Astronomico di Bologna, via Ranzani 1, I-40127 Bologna, Italy
           }


 
  \abstract
   {Studying the coupling between the energy output produced by the central quasar and the host galaxy is fundamental to fully understand galaxy evolution. Quasar feedback is indeed supposed to dramatically affect the galaxy properties by depositing large amounts of energy and momentum into the ISM.}
   {In order to gain further insights on this process, we study the Spectral Energy Distributions (SEDs) of sources at the brightest end of the quasar luminosity function, for which the feedback mechanism is supposed to be at its  maximum, given their high efficiency in driving powerful outflows.}
   {We model the rest-frame UV-to-FIR SEDs of 16 \textbf{WI}SE-\textbf{S}DSS \textbf{S}elected \textbf{H}yper-luminous (WISSH) quasars at $\rm 1.8 < z < 4.6$ based on SDSS, 2MASS, WISE and \textit{Herschel}/SPIRE data. Through an accurate  SED-fitting procedure, we disentangle the different emission components deriving physical parameters of both the nuclear component (i.e. bolometric and monochromatic luminosities) and the host galaxy (i.e. Star Formation Rate SFR, mass and temperature of the cold dust). We also use a radiative transfer code to account for the contribution of the quasar-related emission to the FIR fluxes.} 
   {Most SEDs are well described by a standard combination of accretion disk+torus and cold dust emission. However, about 30\% of them require an additional emission component in the NIR, with temperatures peaking at $\rm \sim 750 \,K$, which indicates the presence of a hotter dust component in these powerful quasars. We measure extreme values of both AGN bolometric luminosity ($ \rm L_{BOL}>10^{47} erg/s$) and SFR (up to $ \rm \sim 2000 \, M_\odot / yr$) based on the host galaxy, quasar-corrected, IR luminosity. A new relation between quasar and star-formation luminosity is derived ( $\rm L_{SF}\propto L_{QSO}^{0.73}$) by combining several \textit{Herschel}-detected quasar samples from $\rm z \sim 0$ to $\rm \sim 4$.
 WISSH quasars have masses ($\rm \sim 10^8 M_{\odot}$)  and temperatures ($\rm \sim 50 \, K$) of cold dust in agreement with those found for other high-$z$ IR luminous quasars.}
 {Thanks to their extreme nuclear and star-formation luminosities, the WISSH quasars are ideal targets to shed light on both the feedback mechanism and its effect on the  evolution of their host galaxies, and the merger-induced scenario which is commonly assumed to explain these exceptional luminosities.\\
 Future observations will be crucial to measure the molecular gas content in these  systems,  probe the impact between quasar-driven outflows and on-going star-formation, and reveal the presence of merger signatures in their host galaxies.}


   \maketitle
%
\section{Introduction}

Whether galaxy formation and supermassive black holes (SMBHs) growth are somehow connected one to each other has been a matter of hot debate in the past decades; if such a connection is real, quasars play a crucial role in the evolutionary scenario. According to the latest models of galaxy-quasar co-evolution, during their active phase the latter are capable to drive energetic outflows from the inner to the outer regions, removing and/or heating up the molecular gas of the host galaxy out of which stars form, thus stopping both star formation and black-hole accretion \citep[see e.g.][]{Silk1998, Hopkins2010,Zubovas2012,Costa2014}. This way, quasars would act as powerful quenching mechanism, preventing massive galaxies to overgrow and being responsible for the transition of their hosts from the blue cloud to the red sequence. \\ 
On the one hand, outflows have been largely observed and studied in the ionised gas component \citep{Holt2008,Nesvadba2008,Rupke2011,Carniani2015,Bischetti2017}, while only few cases are known of powerful outflows involving the molecular gas \citep[]{Feruglio2010,Feruglio2013,Feruglio2015, Sturm2011,Cicone2012,Cicone2014, Aalto2015,Gonzales2017}. It is in fact not yet clear how they are linked.\\

\noindent
On the other hand, some models also require  positive quasar feedback driven by the outflow, in the form of induced star formation in the host galaxy, via gas compression of the interstellar medium \citep[]{King2005,Ishibashi2012,Silk2013}. However, only scanty observational evidences are available \citep{CanoDiaz2012, Cresci2015a, Cresci2015b,Maiolino2017},  supporting this fascinating scenario. 

Since BH-galaxy co-evolution is directly linked to the rate of star formation and consumption of gas, one of the key parameters to study is the star formation (SF) activity of the quasars' host galaxies. This has benefitted from progressively larger numbers of quasars being identified in wide, systematic infrared galaxy surveys. The problem in this type of analysis is that the presence of the quasar may bias the estimate of SF in the host by contaminating the UV and optical range of the spectrum. 
A possible solution might be to use the far-infrared (FIR) emission, due to UV light reprocessed by the dust grains. In fact, the quasar contamination drops towards the FIR band \citep[see, however][]{Schneider2015} and the effect of dust-extinction is almost negligible.

The first investigations of the FIR emission in quasars was carried out with the Infrared Astronomical Satellite (IRAS) and the Infrared Space Observatory (ISO)  combined with data from SCUBA and IRAM thus extending the observational range at longer wavelengths. The launch of the \textit{Herschel} Space Observatory in 2009 \citep{Pilbratt2010} allowed to observe the galaxy rest-frame emission as never before. 
Thanks to \textit{Herschel} observations, several studies have been done in the last years to understand whether the presence of a quasar in the center of the galaxy has any influence on the host Star Formation Rate (SFR). The results have been however contraddictory.  
Most of these studies have found evidence for enhanced SFR in quasars hosts, compared to non-active galaxies of the same stellar mass, and argued that the bulk of moderate-luminosity X-ray selected quasars are hosted in galaxies which trace that of normal star-forming `main-sequence'  (MS) galaxies \citep[e.g.][]{Harrison2012,Mullaney2012a,Rosario2013,Santini2012,Stanley2015}. On the contrary, \citet{Bongiorno2012,Azadi2015,Mullaney2015} reported a broad specific SFR (sSFR)  distribution for X-ray selected quasars peaking below the MS. 

In addition, there is a further analysis to be done to understand if quasars are more likely to reside in quiescent or in star-forming host galaxies, i.e. to look for an overall correlation between the star formation and the nuclear activity in individual galaxies. \citet{Hatziminaoglou2010}, studying a sample of HerMES (\textit{Herschel} Multi-tiered Extragalactic Survey) high-redshift quasars, found the presence of a correlation between the quasar and the SF luminosities for both  high- and low-luminousity sources.  \citet{Shao2010}, alternatively, showed a little dependence on the quasar luminosity for the SF in the latter class, in accordance with the results of \citet{Mullaney2012a} for moderate luminous quasars and of \citet{Rosario2012} who used a larger sample, including the COSMOS one. Conversely, \citet{Mullaney2012b} do find hints of coeval growth of the super-massive black hole and host galaxy, suggesting a causal connection, also found by \citet{Delvecchio2015} in his analysis of about 8500 sources with \textit{Herschel} observations up to $z = 2.5$. 
A possible reason for such discrepant results resides in the fact that there is an additional question to be kept in mind:  how strong is the quasar contamination to the FIR? 
\citet{Symeonidis2017} argued that in their sample of the most powerful quasars collected from \citet{Tsai2015} and \citet{Netzer2016}, the total infrared luminosity is dominated by the quasar emission and there is no need for a star-forming component. On the contrary, \citet{Hatziminaoglou2010} showed that a starburst is always needed to reproduce the FIR emission of their sample. The truth is problably in between: \citet{Haas2003} proposed that objects  hosting the largest FIR luminosities ($\rm L_{FIR} > 10^{13} - 10^{14} L_{\odot}$) have a strong star forming component but the quasar contribution in the FIR is not negligible. Similarly, \citet{Dai2012} and \citet{Schneider2015} concluded that the radiation emitted by the central nucleus can provide an important source of heating for the dust in the galaxy.

In order to understand the discrepant results among the aforementioned works, it is important to constrain any correlation between the quasars and  their host galaxy parameters. Both models and observations have confirmed that the efficiency in driving energetic winds and the momentum fluxes of galaxy-scale outflows increases with the quasar bolometric luminosity \citep[see e.g.][]{Menci2008,Hopkins2016,Fiore2017}. Hence, one would expect that feedback could reach its maximum efficiency at the brightest end of the quasar luminosity function. For this reason, understanding the coupling between the nuclear energy output and the host galaxy is an open issue particularly relevant for the most luminous quasars, especially at 1<z<3, the golden epoch for galaxy-quasar co-evolution.\\

For these reasons, here we present a multi-wavelength analysis of 16 quasars belonging to the wider WISSH (\textbf{WI}SE-\textbf{S}DSS \textbf{S}elected \textbf{H}yper-luminous) sample, which collects about 90 of the most luminous quasars known, with the highest bolometric luminosities, i.e. $\rm L_{BOL} \geq 10^{47} erg/s$.\\ 

\noindent
A detailed Spectral Energy Distribution (SED) analysis has been performed, from UV to FIR, in order to accurately disentangle the different emission components.  
The strength of the SED fitting method is that, given sufficiently wide photometric coverage, it is applicable to all quasars, obscured and unobscured, independently of their luminosity. Moreover, a detailed  radiative transfer model has been used in order to statistically assess the relative contribution to dust heating by the central quasar and the hot stars. 

The paper is organized as follows: In Section \ref{sec:2} we describe the sample and the  data, with focus on the \textit{Herschel} data extraction procedure. Section \ref{sec:3} provides a detailed explanation about the method used to disentangle the different emission components, describing the templates adopted to model the whole spectral energy distribution. The output of the fitting procedure are shown in Section \ref{sec:4}, where we report bolometric and monochromatic luminosities and a study of the infrared properties (host infrared luminosity and SFR, cold and hot dust masses) of the sources. In the same Section we also make a thorough study on the possible contribution of the quasar to the FIR fluxes, gaining a statistical result to be applied to the entire sample. Finally, Section \ref{sec:5} presents the conclusions of our work.

  In what follows, we adopt a $ \rm \Lambda$CDM cosmology with $ \rm H_0 = 67.90 \, km/s/Mpc $, $ \rm \Omega_m = 0.3065$, $\rm \Omega_{\Lambda} = 0.6935$ \citep[][]{Planck2015}.


\section{Sample and Data}
\label{sec:2}
In this paper, we focus on a sub-sample of 16 quasars at $\rm 1.9<z<4.6$ extracted from the WISSH quasar sample. The total WISSH sample \citep[see][hereafter Paper I, for more details]{Bischetti2017} includes 86 quasars obtained by cross-correlating the sources detected at $\rm 22 \, \mu m$  with a flux density $\rm S_{22\, \mu m}>$3 mJy from the Wide-field Infrared Survey Explorer \citep[WISE][]{Wright2010} with the Sloan Digital Sky Survey (SDSS) DR7 optically discovered Type I sources at $\rm z>1.5$.  The mid-IR $\rm 22-\mu m$ selection of the WISSH sample ensures several benefits. In particular, (1) IR-loud sources which could be lost in UV-selected samples are favoured and we tend to select quasars which, as some prominent models predict \citep[see e.g.][]{Sanders1988,Hopkins2006,Hopkins2008}, represent an intermediate population emerging from the merger-driven, heavily-obscured phase and preceding the blue quasar phase in which the gas/dust content in the host galaxy may be already swept away; (2) all quasars can be compared in IR luminosity without the typical uncertainties due to extinction correction affecting the luminosity measurements at shorter wavelengths.\\

\begin{figure}
\includegraphics[scale=0.45]{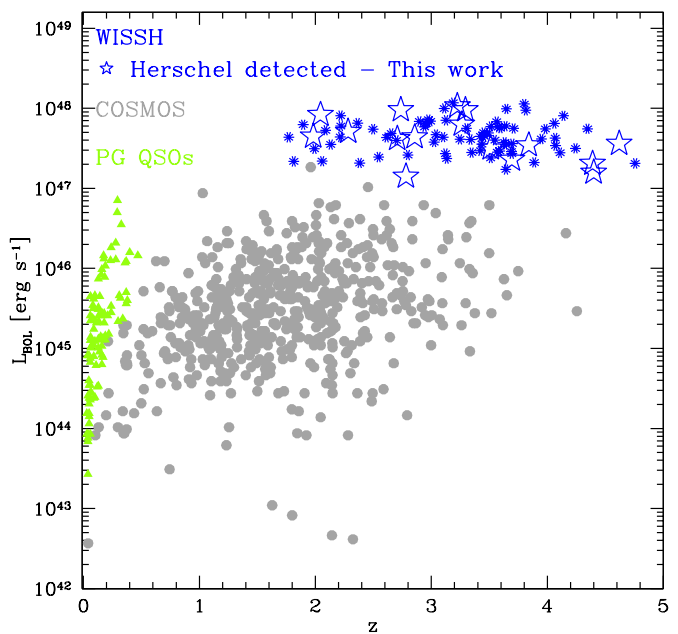}
\caption{Bolometric luminosity versus redshift of the total WISSH sample (in blue). The big stars are the 16 sources with \textit{Herschel} data analyzed in this work. For comparison, also the COSMOS survey by \citep{Bongiorno2012} and the PG QSOs by \citet{Veilleux2009} and \citet{Petric2015} are shown in grey and green respectively.}
\label{fig:lum_z}
\end{figure}

\noindent
WISSH quasars populate the brightest end of the luminosity function, as visibile in Fig. \ref{fig:lum_z}, where we show the $\rm L_{bol}$vs $\rm z$ distribution of the total WISSH sample (in blue), compared to the COSMOS survey by \citep{Bongiorno2012} in grey and the PG quasars by  \citet{Veilleux2009} and \citet{Petric2015} in green.\\
Among the WISSH sources, this work focusses on the 16 quasars for which \textit{Herschel} observations are available on the archive. The lack of a unique preselection in the observational campaigns from which the data have been extracted shall reduce possible selectiones bias. Moreover, as shown in Fig. \ref{fig:lum_z}, the big stars appear to be randomly distributed both in redshift and luminosity. They could be therefore considered representative of the total WISSH sample. For these 16 quasars the bolometric luminosity has been computed as explained in Sec. \ref{sec:bolo} while for the remaining 70 sources we have used the SED-fitting analysis by Duras et al. (in prep.).  \\

\noindent
For all the analyzed sources we have data in the three \textit{Herschel} SPIRE bands (at $\rm 250\, \mu m$, $\rm 350\, \mu m$ and $\rm 500\, \mu m$). They are ideal targets to study host properties as the SFR, IR luminosities and dust masses, and relate them to the nuclear properties, e.g. BH accretion and quasar bolometric luminosities.

In Table \ref{tab:sorgenti} we list the 16 sources studied here, reporting the numeric ID associated to each source, their SDSS ID, and their redshift. Most of the objects have redshift taken from the seventh or the tenth release of the SDSS, but for 5 sources out of 16, we can rely on more accurate z estimates performed by Vietri et al. (in prep.) on LBT/LUCI spectra using the H$\beta$ line.

\begin{table*} 
\begin{center}
\begin{tabular}{cllll}
\toprule\toprule
N & SDSS name & z & Prop \textit{Herschel} ID& Reference \\
\midrule\midrule
\phantom{1}1 & SDSSJ012403.77+004432.6 &   3.840  & $ \rm OT2\_mviero\_2 $ & \citet{Viero2014} \\  
\phantom{1}2 & SDSSJ020950.71-000506.4 &   2.856  & $ \rm OT2\_mviero\_2 $ & \cite{Viero2014} \\
\phantom{1}3 & SDSSJ073502.30+265911.5 &  1.982  & $\rm OT2\_dweedman\_2 $  & \citet{Weedman2016} \\
\phantom{1}4 & SDSSJ074521.78+473436.1 &   3.225 * & $\rm OT2\_dweedman\_2 $  & \citet{Weedman2016} \\	
\phantom{1}5 & SDSSJ080117.79+521034.5 &   3.263 * & $\rm OT2\_dweedman\_2 $ & \citet{Weedman2016} \\	
\phantom{1}6 & SDSSJ081855.77+095848.0 &   3.694 & $\rm OT2\_dweedman\_2 $  & \citet{Weedman2016}   \\	
\phantom{1}7 & SDSSJ090033.50+421547.0 &   3.294 * & $\rm OT2\_dweedman\_2 $  & \citet{Weedman2016}  \\	
\phantom{1}8 & SDSSJ092819.29+534024.1 &   4.390 & $\rm OT2\_dweedman\_2 $  & \citet{Weedman2016} \\	
\phantom{1}9 & SDSSJ101549.00+002020.0 &   4.400 & $\rm OT1\_vkulkarn\_1 $ & Not published \\
10 & SDSSJ121549.81-003432.1 &  2.707 & $\rm KPOT\_seales01\_2 $ & \citet{Eales2010} \\
11 & SDSSJ123714.60+064759.5 &  2.781  & $\rm KPOT\_jdavie01\_1 $ & \citet{Davies2012}\\
12 & SDSSJ125005.72+263107.5 &  2.044  & $\rm KPOT\_seales01\_2 $ & \citet{Eales2010} \\	
13 & SDSSJ143352.20+022713.9 &  4.620  & $\rm OT1\_hnetzer\_2 $ &  \citet{Netzer2014}\\	
14 & SDSSJ170100.60+641209.3 &  2.737 & $\rm OT2\_ymatsuda\_1 $ & \citet{Kato2016}\\
15 & SDSSJ212329.46-005052.9 &  2.283 * & $\rm OT2\_hnetzer\_4 $ & \citet{Netzer2014}\\
16   & SDSSJ234625.66-001600.4 &  3.512 * & $ \rm GT2\_mviero\_1$ & \citet{Viero2014}\\
\bottomrule
\end{tabular}
\end{center}
\caption{Sample of WISSH quasars: source number, SDSS ID, SDSS redshift (* sources have z obtained by Vietri et al. in prep., from LBT NIR spectroscopy or SINFONI spectroscopy) and \textit{Herschel} Proposal IDs (with relative publication where present)}.
\label{tab:sorgenti} 
\end{table*}

\subsection{Multi-wavelength data}
For the 16 sources analyzed here, we have collected data from four different surveys, for a total of 15 bands: SDSS covering the optical bands u, g, r, i and z\footnote{http://skyserver.sdss.org/dr10/en/home.aspx}; 2MASS in the NIR J-, H- and K-bands\footnote{http://irsa.ipac.caltech.edu/frontpage/}, WISE at $\rm 3\, \mu m$, $\rm 4.5\, \mu m$, $\rm 12\, \mu m$ and $\rm 22\, \mu m$ and finally SPIRE in the FIR, at $\rm 250\, \mu m$, $\rm 350\, \mu m$ and $\rm 500\, \mu m$. All the photometric data have been corrected for Galactic extinction \citep{Schlegel1998}.
   
In Table ~\ref{tab:leff}, we report the effective wavelengths of the used filters, while the magnitudes from the UV to the MIR of the sample are shown in Table \ref{tab:photo1} .

\begin{table} 
\begin{center}
\begin{tabular}{c|cc}
\toprule\toprule
Survey/Instrument & Filter & $\rm \lambda_{eff}$\\
\midrule\midrule
\multirow{5}*{SDSS} & u & 3543 $\AA$\\ 
\cline{2-3} & g & 4770 $\AA$\\
\cline{2-3} & r & 6231 $\AA$\\
\cline{2-3} & i & 7625 $\AA$\\ 
\cline{2-3} & z & 9134 $\AA$\\
\cline{1-3}\\
\multirow{3}*{2MASS} & J & 12350 $\AA$\\ 
\cline{2-3} & H & 16620 $\AA$\\
\cline{2-3} & K & 21590 $\AA$\\
\cline{1-3}\\
\multirow{4}*{WISE} & W1 & 3.35 $\mu$m \\
\cline{2-3} & W2 & 4.60 $\mu$m\\
\cline{2-3} & W3 & 11.56 $\mu$m\\
\cline{2-3} & W4 & 22.08 $\mu$m\\
\cline{1-3}\\
\multirow{3}*{\textit{Herschel}} & SPIRE1 &  243 $\mu$m\\
\cline{2-3} & SPIRE2 &  340 $\mu$m\\
\cline{2-3} & SPIRE3 &  482 $\mu$m\\
\bottomrule
\end{tabular}
\end{center}
\caption{Summary of the photometric surveys properties.}
\label{tab:leff} 
\end{table}

\begin{table*}
\begin{center}
\begin{tabular}{ccccccccccccc}
\toprule\toprule
N & u & g & r & i & z & J & H & K & W1 & W2 & W3 & W4 \\
\midrule\midrule
\phantom{1}1 &	$ 23.23	 $ &	$ 19.18  $	& $ 17.86  $ & $	17.92	 $ &	$ 17.82 $ &	$ 17.68  $ & $	17.28	 $ & $ 17.55 $	& $	17.13  $	& $	17.35  $	& $	15.80  $	& $	14.35 $	 \\
\vspace{2mm}
 & \footnotesize{$\pm 0.51$} & \footnotesize{$\pm	0.01$} & \footnotesize{$\pm	0.01$} & \footnotesize{$\pm 0.01$} & \footnotesize{$\pm 0.02$} & \footnotesize{$\pm 0.15$} & \footnotesize{$\pm 0.16$} &  \footnotesize{$\pm 0.22$} & \footnotesize{$\pm 0.03$} & \footnotesize{$\pm 0.04$} & \footnotesize{$\pm 0.09$} & \footnotesize{$\pm 0.17$}  \\
\phantom{1}2 &	$ 18.35	 $ &	$ 17.14  $	& $ 17.05  $ & $	16.93	 $ &	$ 16.81 $ &	$ 16.64  $ & $	16.68	 $ & $ 16.77 $	& $	16.50 $	& $	16.12  $	& $	14.45  $	& $	13.45 $	 \\
\vspace{2mm}
 & \footnotesize{$\pm 0.02$} & \footnotesize{$\pm	0.01$} & \footnotesize{$\pm	0.01$} & \footnotesize{$\pm 0.01$} & \footnotesize{$\pm 0.01$} & \footnotesize{$\pm 0.06$} & \footnotesize{$\pm 0.12$} &  \footnotesize{$\pm 0.13$} & \footnotesize{$\pm 0.02$} & \footnotesize{$\pm 0.03$} & \footnotesize{$\pm 0.09$} & \footnotesize{$\pm 0.07$}  \\
\phantom{1}3 &	$ 16.62	 $ &	$ 16.44  $	& $ 16.22  $ & $	16.03	 $ &	$ 15.89 $ &	$ 15.90  $ & $	15.89	 $ & $ 15.69 $	& $	15.59  $	& $	14.99  $	& $	13.51  $	& $	12.78 $	 \\
\vspace{2mm}
 & \footnotesize{$\pm 0.01$} & \footnotesize{$\pm	0.01$} & \footnotesize{$\pm	0.01$} & \footnotesize{$\pm 0.01$} & \footnotesize{$\pm 0.01$} & \footnotesize{$\pm 0.03$} & \footnotesize{$\pm 0.04$} &  \footnotesize{$\pm 0.04$} & \footnotesize{$\pm 0.02$} & \footnotesize{$\pm 0.02$} & \footnotesize{$\pm 0.02$} & \footnotesize{$\pm 0.06$}  \\
\phantom{1}4 &	$ 19.22	 $ &	$ 16.38  $	& $ 16.17  $ & $	16.15	 $ &	$ 16.09 $ &	$ 15.91  $ & $	15.94	 $ & $ 15.95 $	& $	15.80  $	& $	15.81  $	& $	14.57  $	& $	13.79 $	 \\
 \vspace{2mm}
 & \footnotesize{$\pm 0.03$} & \footnotesize{$\pm	0.01$} & \footnotesize{$\pm	0.01$} & \footnotesize{$\pm 0.01$} & \footnotesize{$\pm 0.01$} & \footnotesize{$\pm 0.04$} & \footnotesize{$\pm 0.05$} &  \footnotesize{$\pm 0.05$} & \footnotesize{$\pm 0.02$} & \footnotesize{$\pm 0.02$} & \footnotesize{$\pm 0.04$} & \footnotesize{$\pm 0.10$}  \\
\phantom{1}5 &	$ 19.49	 $ &	$ 17.08  $	& $ 16.77  $ & $	16.66	 $ &	$ 16.62 $ &	$ 16.58  $ & $	16.70	 $ & $ 16.44 $	& $	16.52  $	& $	16.34 $	& $	14.45  $	& $	13.69 $	 \\
\vspace{2mm}
 & \footnotesize{$\pm 0.03$} & \footnotesize{$\pm	0.01$} & \footnotesize{$\pm	0.01$} & \footnotesize{$\pm 0.01$} & \footnotesize{$\pm 0.01$} & \footnotesize{$\pm 0.07$} & \footnotesize{$\pm 0.13$} &  \footnotesize{$\pm 0.10$} & \footnotesize{$\pm 0.03$} & \footnotesize{$\pm 0.03$} & \footnotesize{$\pm 0.03$} & \footnotesize{$\pm 0.09$}  \\
\phantom{1}6 &	$ 24.98	 $ &	$19.01  $	& $ 17.84 $ & $	17.79	 $ &	$ 17.70 $ &	$ 17.59  $ & $< 17.30	 $ & $ 17.63 $	& $	17.53   $	& $	$17.64	& $	15.49  $	& $	14.40 $	 \\
\vspace{2mm}
 & \footnotesize{$\pm 0.71$} & \footnotesize{$\pm	0.01$} & \footnotesize{$\pm	0.01$} & \footnotesize{$\pm 0.01$} & \footnotesize{$\pm 0.02$} & \footnotesize{$\pm 0.15$} & \footnotesize{$-$} &  \footnotesize{$\pm 0.22$} & \footnotesize{$\pm 0.03$} & \footnotesize{$\pm 0.05$} & \footnotesize{$\pm 0.07$} & \footnotesize{$\pm 0.20$}  \\
\phantom{1}7 &	$ 22.84	 $ &	$17.01  $	& $ 16.68 $ & $	16.63	 $ &	$ 16.54 $ &	$ 16.24  $ & $	16.04	 $ & $ 15.89 $	& $	15.87   $	& $	$15.60	& $	14.74  $	& $	13.89 $	 \\
\vspace{2mm}
 & \footnotesize{$\pm 0.38$} & \footnotesize{$\pm	0.01$} & \footnotesize{$\pm	0.01$} & \footnotesize{$\pm 0.01$} & \footnotesize{$\pm 0.01$} & \footnotesize{$\pm 0.07$} &  \footnotesize{$\pm 0.07$} & \footnotesize{$\pm 0.06$} & \footnotesize{$\pm 0.02$} & \footnotesize{$\pm 0.02$} & \footnotesize{$\pm 0.04$} & \footnotesize{$\pm 0.11$}   \\
\phantom{1}8 &	$ 24.48	 $ &	$23.59  $	& $ 20.28  $ & $	19.46	 $ &	$ 18.53 $ &	$ 18.09  $ & $	17.89	 $ & $ 17.70 $	& $	17.85   $	& $	$17.96	& $	16.24  $	& $	14.72 $	 \\
\vspace{2mm}
& \footnotesize{$\pm 0.92$} & \footnotesize{$\pm	0.32$} & \footnotesize{$\pm	0.03$} & \footnotesize{$\pm 0.02$} & \footnotesize{$\pm 0.03$} & \footnotesize{$\pm 0.21$} &  \footnotesize{$\pm 0.27$} & \footnotesize{$\pm 0.24$} & \footnotesize{$\pm 0.03$} & \footnotesize{$\pm 0.05$} & \footnotesize{$\pm 0.11$} & \footnotesize{$\pm 0.20$}   \\
\phantom{1}9 &	$ 24.12	 $ &	$21.68  $	& $ 19.62 $ & $	19.20	 $ &	$  18.93 $ &	$ -  $ & $	-	 $ & $ - $	& $ 18.20  $	& $ 18.29 $  & $ 16.55 $	& $	15.17 $	 \\
\vspace{2mm}
& \footnotesize{$\pm 0.68 $} & \footnotesize{$\pm	0.06$} & \footnotesize{$\pm	0.02$} & \footnotesize{$\pm 0.02$} & \footnotesize{$\pm 0.04$} & \footnotesize{$ -$} &  \footnotesize{$ -$} & \footnotesize{$ -$} & \footnotesize{$\pm 0.05 $} & \footnotesize{$\pm 0.11$} & \footnotesize{$\pm 0.29$} & \footnotesize{$\pm 0.52$}   \\
10 &	$ 18.96	 $ &	$17.43  $	& $ 17.28 $ & $	17.09	 $ &	$ 16.87  $ &  $	16.69	 $ & $ 16.64 $	& $	16.65  $	& $	$16.49	& $	16.13  $	& $	14.19 $ & $ 13.34  $	 \\
\vspace{2mm}
& \footnotesize{$\pm 0.02$} & \footnotesize{$\pm	0.01$} & \footnotesize{$\pm	0.01$} & \footnotesize{$\pm 0.01$} & \footnotesize{$\pm 0.01$} & \footnotesize{$\pm 0.07$} &  \footnotesize{$\pm 0.07$} & \footnotesize{$\pm 0.12$} & \footnotesize{$\pm 0.03$} & \footnotesize{$\pm 0.03$} & \footnotesize{$\pm 0.04$} & \footnotesize{$\pm 0.09$}   \\
11 &	$ 21.24	 $ &	$ 19.11  $	& $ 18.53 $ & $	18.33	 $ &	$  18.18 $ &	$ -  $ & $	-	 $ & $ - $	& $ 17.33  $	& $ 16.93 $  & $ 14.95 $	& $	13.74 $	 \\
\vspace{2mm}
& \footnotesize{$\pm 0.10 $} & \footnotesize{$\pm	0.01$} & \footnotesize{$\pm	0.01$} & \footnotesize{$\pm 0.01$} & \footnotesize{$\pm 0.02$} & \footnotesize{$ -$} &  \footnotesize{$ -$} & \footnotesize{$ -$} & \footnotesize{$\pm 0.03 $} & \footnotesize{$\pm 0.04$} & \footnotesize{$\pm 0.05$} & \footnotesize{$\pm 0.11$}   \\
12 &	$ 15.52	 $ &	$15.58  $	& $ 15.52 $ & $	15.34	 $ &	$ 15.19 $ &  $	15.24	 $ & $ 15.22 $	& $	15.03  $	& $	$15.17	& $	14.77 $	& $	13.36 $ & $ 12.68  $	 \\
\vspace{2mm}
& \footnotesize{$\pm 0.01$} & \footnotesize{$\pm	0.01$} & \footnotesize{$\pm	0.01$} & \footnotesize{$\pm 0.01$} & \footnotesize{$\pm 0.01$} & \footnotesize{$\pm 0.04$} &  \footnotesize{$\pm 0.05$} & \footnotesize{$\pm 0.04$} & \footnotesize{$\pm 0.02$} & \footnotesize{$\pm 0.02$} & \footnotesize{$\pm 0.02$} & \footnotesize{$\pm 0.05$}   \\
13 &	$ 22.64	 $ &	$ 21.61  $	& $  19.42 $ & $	18.25	 $ &	$  18.18 $ &  $	17.43	 $ & $ 17.56 $	& $	17.43  $	& $	$ 17.78	& $	17.61 $	& $	15.89 $ & $ 14.58  $	 \\
\vspace{2mm}
& \footnotesize{$\pm 0.43$} & \footnotesize{$\pm	0.07$} & \footnotesize{$\pm	0.02$} & \footnotesize{$\pm 0.01$} & \footnotesize{$\pm 0.03$} & \footnotesize{$\pm 0.16$} &  \footnotesize{$\pm 0.01$} & \footnotesize{$\pm 0.22$} & \footnotesize{$\pm 0.03$} & \footnotesize{$\pm 0.05$} & \footnotesize{$\pm 0.08$} & \footnotesize{$\pm 0.18$}   \\
14 &	$ 16.61	 $ &	$15.96  $	&  $	15.87	 $ &	$ 15.79 $ &  $	15.73	 $ & $ 15.64 $	& $	15.68  $	& $	$15.77	& $	15.70 $	& $	15.49 $ & $ 13.88  $	  & $ 13.09  $\\
\vspace{2mm}
& \footnotesize{$\pm 0.01$} & \footnotesize{$\pm	0.01$} & \footnotesize{$\pm	0.01$} & \footnotesize{$\pm 0.01$} & \footnotesize{$\pm 0.01$} & \footnotesize{$\pm 0.05$} &  \footnotesize{$\pm 0.04$} & \footnotesize{$\pm 0.05$} & \footnotesize{$\pm 0.02$} & \footnotesize{$\pm 0.02$} & \footnotesize{$\pm 0.02$} & \footnotesize{$\pm 0.04$}   \\
15 &	$ 16.94	 $ &	$16.47  $	&  $	16.30	 $ &	$ 16.25 $ &  $	16.04	 $ & $ 16.05 $	& $	15.98  $	& $	$15.74	& $	16.01 $	& $	15.51 $ & $ 13.95   $	  & $ 13.11  $\\
\vspace{2mm}
& \footnotesize{$\pm 0.01$} & \footnotesize{$\pm	0.01$} & \footnotesize{$\pm	0.01$} & \footnotesize{$\pm 0.01$} & \footnotesize{$\pm 0.01$} & \footnotesize{$\pm 0.05$} &  \footnotesize{$\pm 0.07$} & \footnotesize{$\pm 0.06$} & \footnotesize{$\pm 0.02$} & \footnotesize{$\pm 0.02$} & \footnotesize{$\pm 0.01$} & \footnotesize{$\pm 0.06$}   \\
16 &	$ 21.03	 $ &	$18.53  $	&  $	17.79	 $ &	$ 17.67 $ &  $	17.50	 $ & $ 17.74 $	& $	17.50  $	& $	$17.14	& $	17.25 $	& $	17.26 $ & $ 15.59  $	  & $ 14.26  $\\
\vspace{2mm}
& \footnotesize{$\pm 0.07$} & \footnotesize{$\pm	0.01$} & \footnotesize{$\pm	0.01$} & \footnotesize{$\pm 0.01$} & \footnotesize{$\pm 0.01$} & \footnotesize{$\pm 0.16$} &  \footnotesize{$\pm 0.18$} & \footnotesize{$\pm 0.16$} & \footnotesize{$\pm 0.03$} & \footnotesize{$\pm 0.05$} & \footnotesize{$\pm 0.08$} & \footnotesize{$\pm 0.17$}   \\

\bottomrule
\end{tabular}
\end{center}
\caption{UV to MIR photometry available for the analyzed sample. Magnitudes are in AB system.}
\label{tab:photo1} 
\end{table*}

While for the UV-to-NIR data we have collected the data from published catalogs, the photometry in the FIR bands has been extracted from archival data. 

\subsubsection{\textit{Herschel} SPIRE data: source extraction}
For the FIR photometry we have used \textit{Herschel} archival data as specified in Table \ref{tab:sorgenti}. For each source we have estimated the \textit{Herschel} flux densities at SPIRE wavelengths using the {\tt sourceExtractorTimeline} \citep{Bendo2013}, a fitting method implemented in HIPE \citep{Ott2010}, and widely used for point-like sources \citep[see for example][]{Pappalardo2016, Ciesla2012}. 
\textit{Herschel} observations are performed scanning ortogonally each target with the SPIRE bolometers, measuring the fluxes at regular time intervals. The measurements set of the bolometers for each cross-scan is called "timeline" data. The {\tt sourceExtractorTimeline} fits these timeline data from all bolometers within  an  individual  array  with  a  two-dimensional  Gaussian function. The input are the source position, a radius containing the source, and an aperture identifying the background annulus.  Following the prescriptions of \citet{Pappalardo2015} we set a search radius for the target of 22", 30", and 42", at $\rm 250\, \mu m$, $\rm 350\, \mu m$ and $\rm 500\, \mu m$, respectively. For the background annulus we define a set of different apertures between 140" and 220", and we estimate the background from the median of all the values recovered.

According to \citet{Pappalardo2015}, two fits are performed to the timeline data. The first one is performed using a circular Gaussian function in which the FWHM is allowed to vary. Then the FWHM is used to determine whether the source is resolved or unresolved and reject sources that are either too narrow (which may be unremoved glitches) or too broad (which are probably extended sources). To test at which distance two sources are distinguishable, point-like objects with the same flux density and an increasing distance from each other are injected in the timeline data. What we found is that two sources are resolved only if they are at a distance above $\sim$ 22", 30" and 46" at 250, 350, and 500 $\rm \mu  m$, respectively. Fixing the distance between the sources to these values, as second step point-like objects at different flux densities are injected in the timeline data, to investigate the FWHM recovered by sourceExtractorTimeline. These tests demonstrate that artificial sources with FWHM corresponding to the telescope beam (17.5", 23.9" and 35.1" at 250, 350, and 500 $\rm \mu  m$, respectively) added to timeline data may have FWHM between 10" and 30" at 250 $\rm \mu  m$, 13.3" and 40" at 350 $\rm \mu  m$, and 20" and 60" at 500 $\rm \mu  m$. Therefore, sources, whose FWHM are determined via sourceExtractorTimeline and were outside these values have been excluded from the analysis. 

Although other  software  packages  have  been  developed  for source extraction within confused extragalactic fields (including software  developed  specifically  for \textit{Herschel}),  we prefer  to use the timeline fitter because the SPIRE data are flux calibrated at the timeline level using timeline-based PSF-fitting techniques,  so  the  method  that  we  have used is consistent with the flux calibration measurements themselves and is therefore expected to be more accurate.\\
In Table \ref{tab:herdata} we report the photometric flux densities obtained from the images.

\begin{table} 
\begin{tabular}{cllll}
\toprule\toprule
N & $\rm f_{250 \, \mu m} [mJy] $ & $\rm f_{350 \, \mu m} [mJy] $ & $\rm f_{500 \, \mu m} [mJy] $  \\
\midrule\midrule
\phantom{1}1 & < 48  & $ < 44 $ & $ < 40 $\\
\phantom{1}2  &  \phantom{1} 71 $ \pm $ 18  &  \phantom{1} 61 $ \pm 10 $ &  \phantom{1} 31 $ \pm 10 $ \\
\phantom{1}3 &  \phantom{1} 91 $ \pm  $ 7.2 &  \phantom{1} 53 $ \pm 7.3 $ & \phantom{1} 33 $ \pm 9.2 $ \\
\phantom{1}4 &  \phantom{1} $55 \pm 7.4 $ &  \phantom{1} 58 $ \pm 7.3 $ &   \phantom{1} 52 $ \pm 9.5 $ \\
\phantom{1}5 &  \phantom{1} $93 \pm 7.3 $ &  \phantom{1} 80 $ \pm 8.1 $ &   \phantom{1} 57 $ \pm 10 $\\
\phantom{1}6 &  \phantom{1} $48 \pm 7.2 $ &  \phantom{1} 49 $ \pm 7.3 $ & $ < 56 $ \\
\phantom{1}7 &  \phantom{1} $29 \pm 7.7 $ &  \phantom{1} 19 $ \pm  7.4 $ & $ < 27 $\\
\phantom{1}8 &  \phantom{1} $65 \pm 7.7 $ &  \phantom{1} 75 $ \pm 8.2 $ &  \phantom{1} 66 $ \pm 9.8$ \\
\phantom{1}9 &  \phantom{1} $20 \pm 5.2 $ &  \phantom{1} 23 $ \pm 5.1 $ &  \phantom{1} 19 $ \pm 6.6$ \\
10 &   \phantom{1} $75 \pm 12 $ &  \phantom{1} 58 $ \pm 12 $ &  \phantom{1} 51 $ \pm 8.3 $ \\
11 &  \phantom{1} $93 \pm 5.1 $ &  \phantom{1} 94 $ \pm 5.1 $ &  \phantom{1} 62 $ \pm 6.4 $\\
12 &  $< 45 $ & $  < 31 $ & $ <  65 $\\
13 &  \phantom{1} $28 \pm 2.9 $ &  \phantom{1} 23 $ \pm 2.9 $ & \phantom{1} 16 $ \pm 3.6 $\\
14 &  \phantom{1} $81 \pm 3.0 $ &  \phantom{1} 56 $ \pm 2.9 $ & $ < 40 $\\
15 &  \phantom{1} $36 \pm 3.7 $ &  \phantom{1} 28 $ \pm 3.6 $ & \phantom{1} 18 $ \pm 4.5 $\\ 
16 &  \phantom{1} $41 \pm 14 $ &   \phantom{1} 38 $ \pm 9.9 $ & $ < 35 $\\
\bottomrule
\end{tabular}
\caption{FIR flux densities in the three \textit{Herschel}-SPIRE bands for the WISSH quasars.}
\label{tab:herdata} 
\end{table}

\section{UV-to-FIR Spectral Energy Distribution}
\label{sec:3}

In this section, we describe the tool we developed to study the Spectral Energy Distribution (SED) of our sources. The approach is based on the idea that the overall observed SED of each quasar is  the result of the combination of different components which can be disantangled, i.e. the nuclear emission and the galactic stellar light. 
The nuclear emission is characterized by two bumps in the UV and NIR regimes \citep{Sanders1989, Elvis1994, Richards2006} which create a dip at around 1$\mu$m. The UV bump is due to thermal emission from the accretion disk \citep{Czerny1987}, while the NIR bump is due to the intrinsic primary radiation absorbed by hot dusty clouds in the torus and subsequently re-emitted at longer wavelength. 
In the same way, the galactic stellar light is made of a combination of direct UV emission by hot stars and its re-emission in the FIR due to the reprocessed light by the cold dust component present at galactic scales. These components will be described in more detail in Section \ref{sec:components}.

\subsection{Modelling the whole SED using a multi-components fitting procedure}

The approach we have followed is based on a three-components fitting procedure 
based on the weighted least square method, better known as the standard reduced $\rm \chi^2$ minimization, where the reduced $\rm \chi^2$ is in the form:

\begin{equation}
\chi^2_{\nu} = \frac{1}{n-p} \sum_{i=1}^n \left ( \frac{f_i^{obs} - f_i^{model}}{\sigma_i} \right )^2
\end{equation}

\noindent
with n: observed data points, $\rm f_i^{obs}$ and $\rm \sigma_i$: observed fluxes and associated errors of the photometric bands,  $\rm f_i^{model}$ the total flux of the chosen model with $\rm p$ free parameters, and $\rm \nu = n - p$ the degrees of freedom. \\
Our model ($\rm f_{model}$) includes three emission components: 
\begin{equation}
f_{model} = A \,\, f_{AD+T} + B \,\, f_{CD} + C \,\, f_{NE} 
\end{equation}

\noindent
where $f_{\textbf{AD+T}}$ represents the emission coming from the \textbf{A}ccretion \textbf{D}isk (both direct and reprocessed by the dusty \textbf{T}orus), $f_{\textbf{CD}}$  accounts for the FIR emission due to the reprocessed flux by \textbf{C}old \textbf{D}ust and is modelled as a modified blackbody,  $f_{\textbf{NE}}$ is any possible \textbf{N}IR emission \textbf{E}xcess modelled with a simple blackbody. Finally, A, B and C are the relative normalizations. For each component we have created a library of templates described below. 
The fitting procedure, through $\chi^2$ minimization, allows us to determine the combination of templates which best describes the observed SED and their relative contribution.


\subsection{Library templates}
\label{sec:components}
The data have been fitted with three main emission components (see previous section). For each component, a number of templates spanning a large grid of physical parameters, have been created. In the next sections we will describe them in detail.

\subsubsection{Quasar emission}
The first component of the SED-fitting tool includes the direct emission from the central engine and the light absorbed and re-emitted by the dusty torus. \\

\noindent
\textbf{(i)   Nuclear Emission}\\



The shape of the SED of the primary source is commonly described by power laws of different spectral indices which however vary considerably from one work to another. 
In order to evaluate the best shape to describe the observed emission of a generic quasar, we have analyzed different models of UV emission used in literature. 

In the work by \citet{Fritz2006}, renewed by \citet{Stalevski2012,Stalevski2016}, the primary source of emission has been modelled as a composition of power laws with indices from \citet{Granato1994} and \citet{Nenkova2002}, i.e. 

\begin{equation}
\lambda L_{\lambda}=\begin{cases} \lambda^{1.2} & 0.001 \leq \lambda \leq 0.030\;[\mu m] \\ \lambda^{0} & 0.030 < \lambda \leq 0.125\;\,[\mu m]\\
\lambda^{-0.5} & 0.125 < \lambda \leq 20.0 \,\;\;\,[\mu m] \, .
\end{cases}
\end{equation}




In 2012 \citet{Feltre2012} proposed an updated version of this model, changing the spectral indices of the power laws according to \citet{Schartmann2005}, making them steeper at small wavelengths:

\begin{equation}
\lambda L_{\lambda}=\begin{cases} \lambda^{2} & 0.001 \leq \lambda \leq 0.050\,\;[\mu m] \\ \lambda^{1.8} & 0.050 < \lambda \leq 0.125\;\,[\mu m]\\
\lambda^{-0.5} & 0.125 < \lambda \leq 10.0 \,\;\;\,[\mu m]\\
\lambda^{-3} &  \lambda > 10.0 \,\;\;\;\;\;\;\;\;\;\;\;\;\;\;\; [\mu m] \, .
\end{cases}
\end{equation}

\begin{figure}
\includegraphics[scale=0.45]{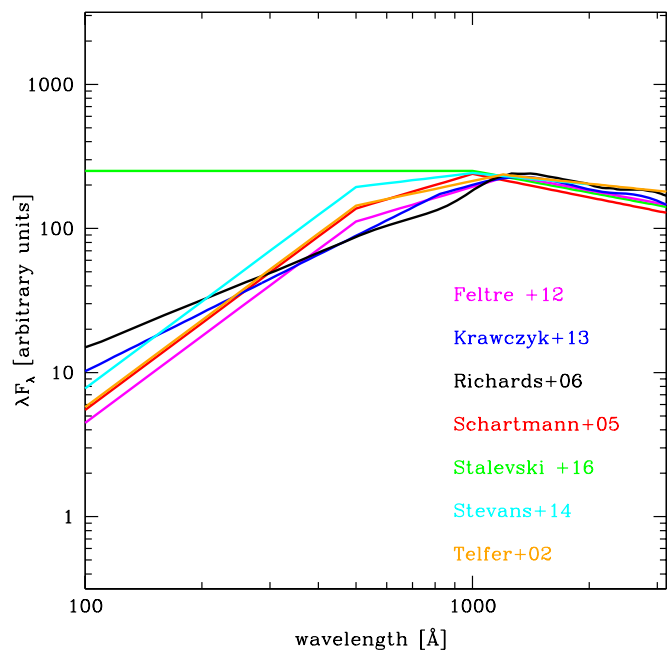}
\caption{Comparison among the SED for different emission models of the quasar primary source.}
\label{fig:bpl}
\end{figure}

Figure \ref{fig:bpl} shows the aforementioned models from \citet{Stalevski2012} and \citet{Feltre2012} compared to the models derived from the composite quasar spectra by \citet{Telfer2002} and \citet{Stevans2014} based on HST quasars spectra, and to the empirical SDSS quasar mean spectra from \citet{Richards2006} and \citet{Krawczyk2013}. As visible in Figure \ref{fig:bpl}, the slope of Stalevski's power-law at short wavelengths ($\lambda<$\, 0.125\, $\mu$m) seems to be at odds with all the other emission models, both empirical and theoretical ones. In our library of templates, we chose therefore to model the accretion disk emission using the recipe by \citet{Feltre2012}. \\

\noindent
\textbf{(i)   Torus Emission}\\

Several models for the radiation emitted by the dusty torus surrounding the accretion disk are available in the literature and they often result in very different solutions when fitted to the data. \citet{Feltre2012} studied the differences between smooth and clumpy dusty torus by comparing the works by \citet{Fritz2006} and \citet{Nenkova2008} and concluded that the models with only a smooth grain distribution are not a realistic description of the torus, since they imply a large number of collisions which would raise the temperature so high to destroy the dust grains \citep[see also][]{Krolik1988}. However, due to the difficulties in handling clumpy media and the lack of computational power, smooth models were the first to be used.
Using a Monte Carlo radiative transfer code (SKIRT), \citet{Stalevski2012,Stalevski2016} simulated a dusty torus as a 3D two-phase medium made of a combination of high density clumps and low density medium filling the space between the clumps. The models have been obtained starting from the smooth models by \citet{Fritz2006} and applying the algorithm described by \citet{Witt1996} to generate a two-phase clumpy medium, according to which each individual cell in the grid is assigned randomly, by a Monte Carlo process, to either a high- or a low-density state.
In such models, the dust is a mixture of silicate and graphite grains \citep[as in][]{Mathis1977} and with optical properties from \citet{Laor1993} and \citet{Li2001}.
Its distribution is described by a flared disc whose geometry is defined by the inner ($\rm R_{in}$) and the outer ($\rm R_{out}$) radii, and by the half-opening angle $\rm OA$, which measures the dust-filled zone, from the equator to the edge of the torus, which is linked to the covering factor.


Concerning other models of dusty tori, we considered the one by \citet{Honig2010}, which is similar to Stalevski's one in making use of a three-dimensional transfer code based on a Monte Carlo approach. However, the model proposed by \citet{Stalevski2012,Stalevski2016} provides the best description of the reprocessed emission by the torus, being both smooth and clumpy, and therefore it represents the best choice to constrain with high accuracy the IR part of the spectrum of our sources. 

\begin{figure}
\begin{center}
\includegraphics[scale=0.45]{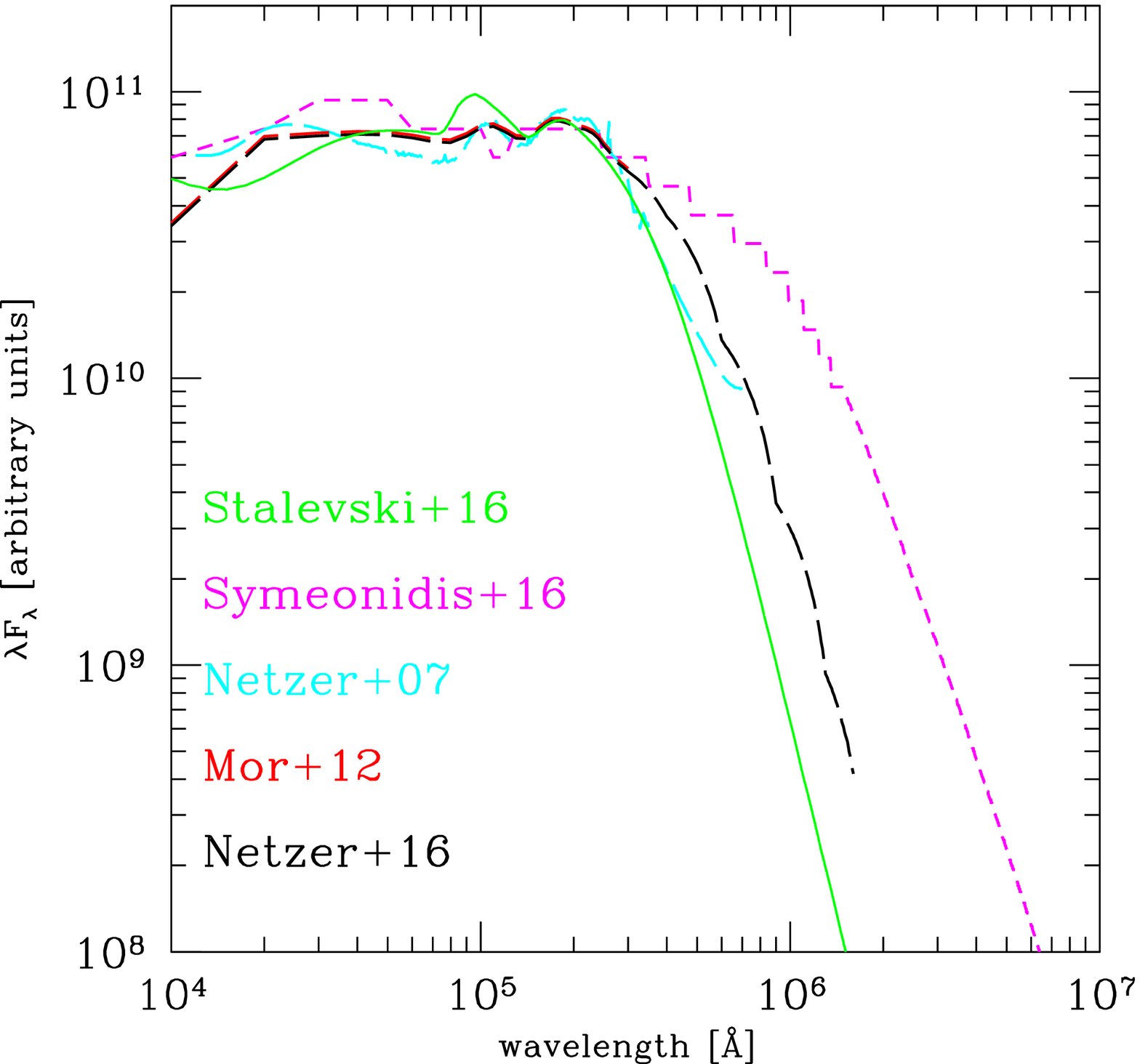}
\caption{Infrared torus model by \citet{Stalevski2016} compared with the empirical SEDs by \citet{Netzer2007},\citet{Mor2012}, \citet{Netzer2016} and \citet{Symeonidis2016}.}
\label{fig:torusmodels}
\end{center}
\end{figure}

Among the available empirical infrared quasar SEDs, we have compared our choice of the torus emission with the SEDs proposed by \citet{Netzer2007}, \citet{Mor2012}, \citet{Netzer2016} \citep[an extended version of][template at longer wavelenghts]{Mor2012}, and \citet{Symeonidis2016}; these are shown in Fig. \ref{fig:torusmodels} in cyan, red, black and magenta respectively. Except for the SED proposed by \citet{Symeonidis2016}, our torus model (in green) is in good agreement with all the others. As more detailed in Sec. \ref{sec:sfr}, we found values of SFR consistent with those published by \citet{Netzer2016} for three sources in common between the two samples, although different tori templates have been adopted. Focusing on the longest wavelengths, the empirical SED by \citet{Symeonidis2016} shows a much higher contribution (lower temperature) compared to the others. This is probably due to the fact that in \citet{Symeonidis2016} the torus extent is larger than in the other cases. From a theoretical point of view, a huge radius of the torus \citep[as the one described by][]{Symeonidis2016} should predict an height of the torus itself difficult to be supported. Moreover, as recently pointed out by \citet{Lani2017}, the discrepancy found between the mean SED by \citet{Symeonidis2016} and the other tori templates shown, might be due to the fact that the first has been obtained without normalising the individual SEDs at a given wavelength and, above all, that it does not represent the entire quasar population but rather the extremely IR-luminous ones. \\

In view of the above, for this work we have created new templates accounting for the accretion disk plus torus emission, made by the combination of the just described quasar torus emission from \citet{Stalevski2016}, and the nuclear emission from \citet{Feltre2012}, appropriately normalized to preserve the energy balance between the UV and the IR bands.

\begin{figure*}
\begin{center}
\includegraphics[scale=0.65]{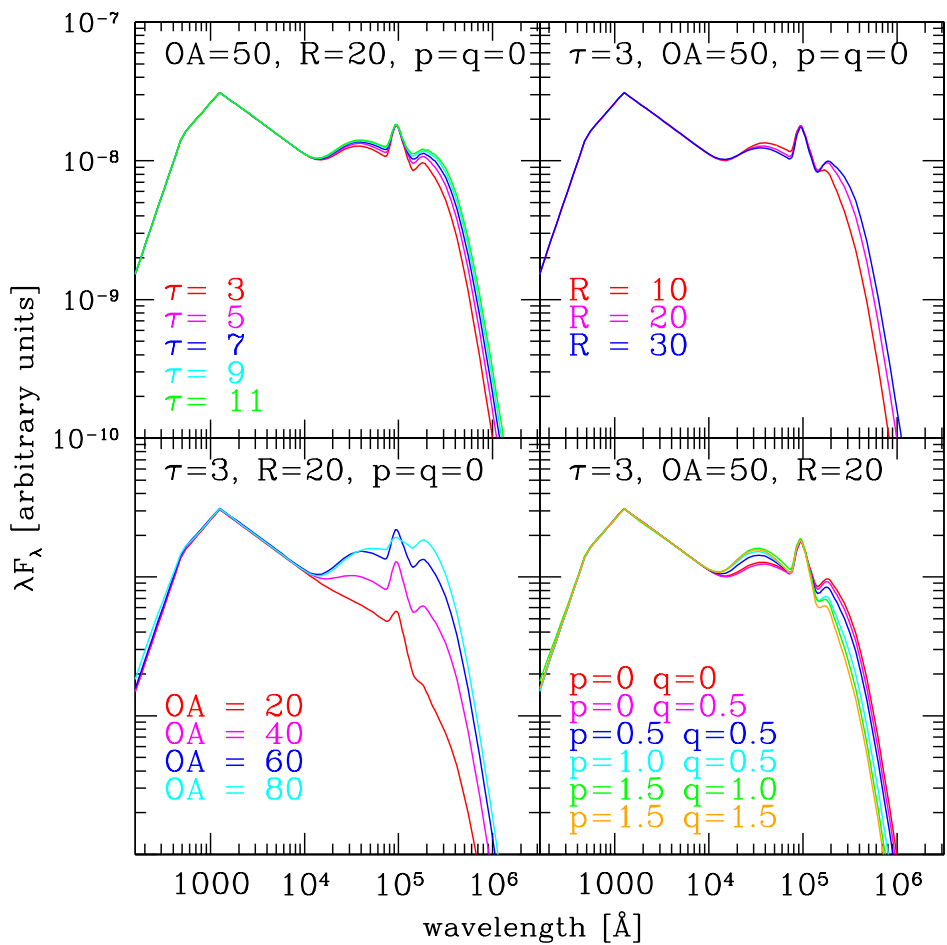}
\caption{Examples of templates at inclination angle $\rm i = 0^{\circ}$ used to describe the AD+T emission.  Different panels show how the torus emission changes varying the physical parameters, i.e. optical depth $\rm \tau$, ratio of outer to inner radius $\rm R=R_{out}/R_{in}$, half opening angle OA, and the $\rm p$ and $\rm q$ parameters related to the density gradient along the radial direction and with polar angle $\rm \rho(r,\theta)\propto r^{-p} e^{-q|cos\theta|}$ \citep{Granato1994}.}
\label{fig:torustempl}
\end{center}
\end{figure*}

Such library is made of 1920 templates with different values of optical depth $\rm \tau$ at $\rm 9.7\, \mu m$ , half-opening angle $\rm OA$ and dust distribution. In addition, for each of them there are ten different lines of sight $\rm \theta$, from face-on ($0^{\circ}$, for typical unobscured type I AGN) to edge-on  ($\rm 90^{\circ}$, obscured type II AGN) view, for a total of 19200 templates (in Figure \ref{fig:torustempl} torus templates for different varying parameters are shown). 
However, in order to reduce the computational time, we used only those with inclination angle $\rm 0^{\circ}$, which is a reasonable assumption considering that we are dealing with type I AGN and that the inclination angle $i$ and the optical depth $\tau$ are strongly degenerate.
In this way, we got a reduced library of 1920 accretion disk plus torus templates.

\subsubsection{Cold Dust Emission}

The second component of the SED-fitting procedure is a modified blackbody accounting for the emission powered by star formation, which is  absorbed by dust grains and re-emitted in the MIR and FIR.  We know that the stellar light is emitted in the UV-to-NIR bands of the spectrum, with the NIR dominated by the older stars, while the UV by the massive, short-lived stars. The same stars, are responsible for polluting the ISM with metals and dust, producing grains which 
absorb the stellar light and re-radiate it in the IR and sub-mm domains.\\

\noindent
In the far-infrared, this dust emits thermal emission characterized by a blackbody spectrum with an additional $\rm \lambda^{-\beta}$ term which accounts for the emissivity of the dust \citep[]{Hildebrand1983}. It must be stressed that since the FIR emission is produced by a mixture of grains of different shapes and sizes, this comes out in a distribution of temperatures. For simplicity one can assume a single-temperature modified blackbody emission even if the SEDs of real galaxies are obviously more complicated.
Following \citet{Blain2002} the rest-frame emission of this cold dust is modeled as:

\begin{equation}
MBB(T_{CD},\lambda) = \lambda^{-\beta} \, BB(T_{CD},\lambda)
\end{equation}
where $\rm BB(T_{CD},\lambda)$ is the Planck function for a dust temperature $\rm T_{CD}$ and $\beta$ is the emissivity index.
The temperature of the modified blackbody and the normalization are free parameters.

The emissivity of dust grains is generally taken to be a power-law at these long wavelengths, with models and laboratory data suggesting indices ranging from $1 \le \beta \le 2$. One of the major sources of uncertainty relies in the exact determination of this value: using different values of $\rm \beta$ produces a systematic change in the best-model value of $\rm T_{CD}$ since a smaller $\rm \beta$ implies a larger cold dust temperature.
Following \citet{Beelen2006}, we fixed $\rm \beta$ to $1.6$ which seems to be the most appropriate value  for high-z quasars. 
Recently, works based on Planck and \textit{Herschel} data have shown that the value of $\rm \beta = 1.6$ (instead $\rm \beta = 2.0$ used in the past) might be the most correct one to describe also the dust in local galaxies; see for example the studies on the Milky Way by \citet{Bianchi2017} and the Section 4.3 of \citet{Planck2014}.

The cold dust library consists of about 50 BB templates, with temperatures in the range $\rm 20 \,K < T < 70 \,K$.


\subsubsection{NIR Emission Excess}
\label{sec:nirexcess}

\begin{figure*}
\begin{center}
\includegraphics[scale=0.8]{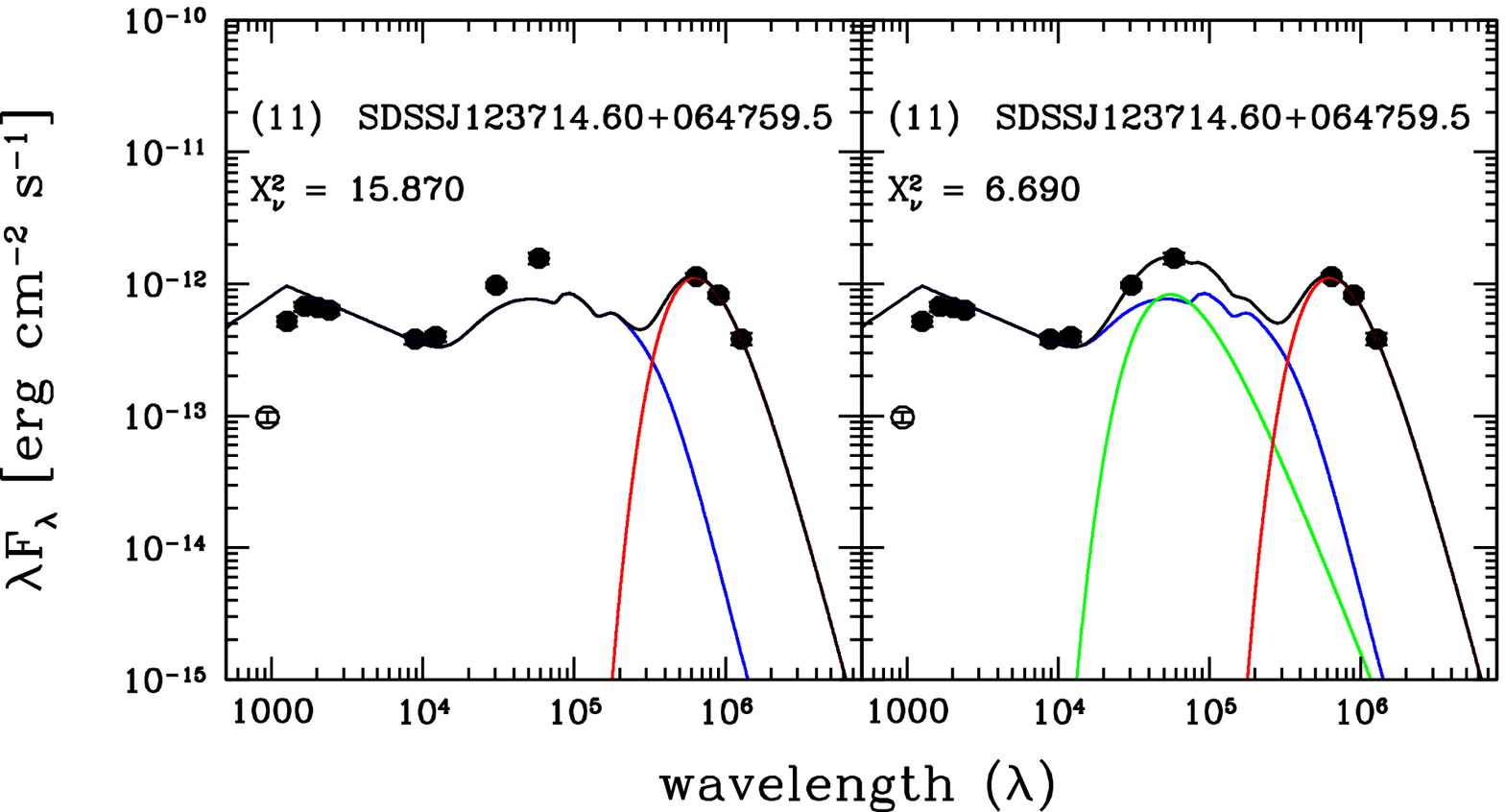}
\caption{Example of a fit in which an additional component (NE) has been added to fully describe the NIR emission. The left panel shows the standard fit with two components while the right panel with the NIR excess blackbody  included. The F-test on this source tells us that such component improves the fit at $\geq$99\% confidence level.}
\label{fig:bbNIRsino}
\end{center}
\end{figure*}

As outlined by several authors in the past, and shown in Figure \ref{fig:bbNIRsino}, an additional component in the near-infrared is often required to fit the NIR part of the SED of luminous quasars \citep[see][]{Barvainis1987, Mor2009, Leipski2013}. \\
In the works mentioned above the authors speculated that the dust grains are probably made of pure-graphite since they are able to survive to temperatures higher than the silicate ones present in the torus \citep{Minezaki2004,Suganuma2006}. The physical scale of such hot-dust is delimited by the sublimation radius of  “typical” dust composed of both silicate and graphite grains (torus inner radius) and the sublimation radius of pure-graphite grains (which has a sublimation temperature of $\rm \sim 1500 \,K$).  

To describe the NIR emission excess, we have generated about $650$ templates using blackbodies of different temperatures, from $\rm T_{NE}=190 \,K $ to $\rm T_{NE}=1800 \,K$. \\

\begin{figure*}
\begin{center}
\caption{Resulting rest-frame SED decompositions for half of the sources analyzed (continues in Figure \ref{fig:sed2}). Black circles are the rest-frame photometric points corresponding to the observed bands used to constrain the SED. Black open circles represent the photometric points not included in the fits (at $\lambda<1216\, \AA$ due to Ly-$\alpha$ absorption, see the text for more details), while the black arrows correspond to upper limits on the observed flux densities at $ 1 \sigma$. Lines  (and shaded areas) correspond to the model templates (and the $1\sigma$ error) found as best-fit solution to describe the photometric points through the $\chi^2$ minimization: blue is the accretion disk plus torus template (AD+T), red is the Cold dust component (CD) while dark green the NIR excess (NE) when present. Finally, the black line shows the sum off all these contributions. Blue, orange and light green shaded areas correspond to the accretion disk plus torus, the cold dust and the NIR excess templates within $1\sigma$ of the best-fit template, and light grey shaded area to their sum.}
\includegraphics[scale=0.92]{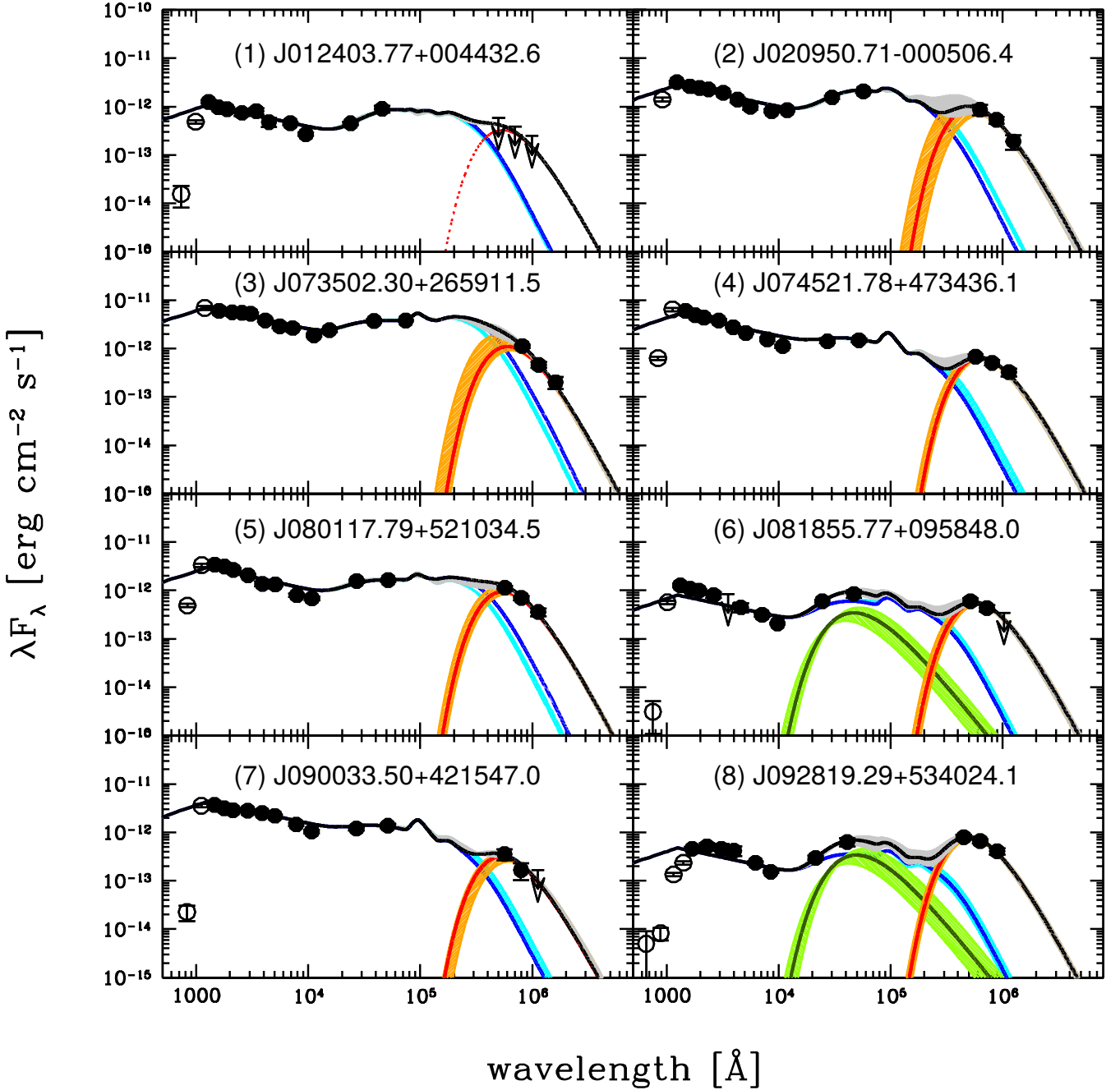}
\label{fig:sed1}
\end{center}
\end{figure*}

\begin{figure*}
\begin{center}
\caption{Same as in Figure \ref{fig:sed1} for the remaining 8 sources.}
\includegraphics[scale=0.98]{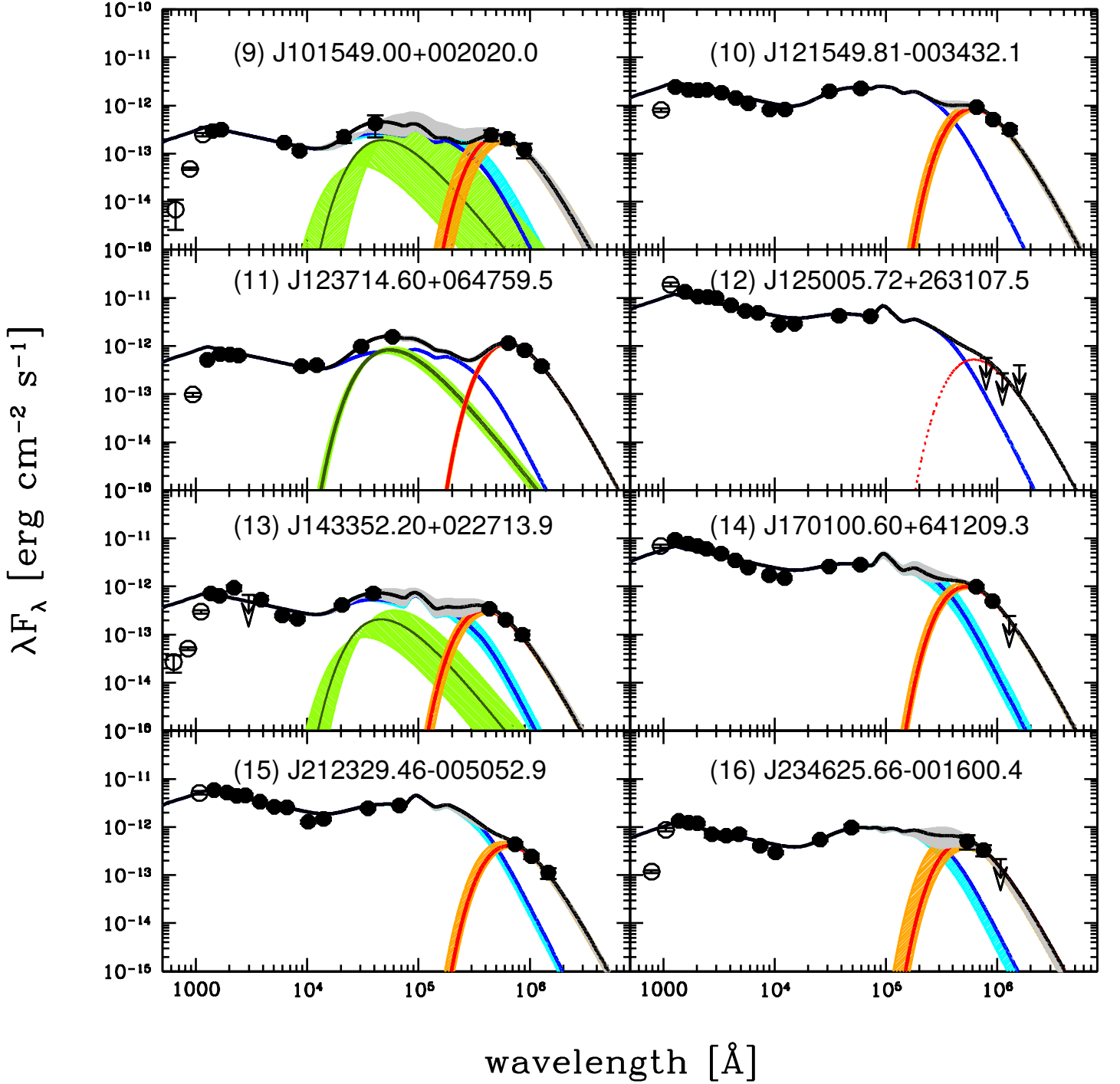}
\label{fig:sed2}
\end{center}
\end{figure*}

\section{Results of the SED fitting}
\label{sec:4}
Given the wide multi-wavelength coverage, the fitting technique described above allows us to decompose the entire spectral energy distribution into the different emission components and to derive robust measurements of both the quasar and the host galaxy properties.
In Figure \ref{fig:sed1} and \ref{fig:sed2} we show the best fit SED model for the 16 sources analyzed here. Black circles are the rest-frame photometric points corresponding to the observed bands used to constrain the SED. Data below $\rm 1216 \, \AA$ corresponding to the Lyman-$\alpha$ emission line, are plotted as open circles and are not taken into account in the fit. At such wavelengths, the flux is expected to be weakened due to the Lyman $\alpha$ absorption forest, not included in our model.  Data points above $\rm 1216 \, \AA$ have been fitted with a combination of the components described above. In detail: (i) accretion disk + torus emission (AD+T, blue line with cyan shaded area); (ii) cold dust emission in the IR band (CD, red line with orange shaded area);  and (iii) NIR excess when necessary (NE, green line with shaded area). The shaded areas describe the range of values corresponding to the solutions for which $\rm \Delta\chi^2 = \chi^2(sol)-\chi^2(best) \le 1.0$, $1\sigma$ in the case of one parameter of interest \citep[see][]{Avni1976}. 

As a first step we have considered in the fit a standard combination of AD+T and CD components. However, in few cases we found that this models are  not enough to reproduce the whole SED, which shows an excess of emission in the NIR bands. For these cases, we have therefore  added as a third component, the  additional NIR excess template, while fixing the AD+T and CD best emission models found in the first run.

To quantify whether such a component really does improve the fit, we have performed the F-test, which measures the goodness of two nested models through the comparison of the chisquares and degrees of freedom obtained in the two cases.

Figure \ref{fig:bbNIRsino} shows as an example, one of the sources for which the additional component has been added (SDSSJ123714.60+064759.5). As visible even by eye, the inclusion of the NIR excess component improves the fit, i.e. looking at the WISE 3 photometric point, we found that including the NIR excess component brings the fit a factor of $\sim$5 closer to the observed point. 
For SDSSJ123714.60+064759.5, the result of the F-test is F=16.67, meaning that the addition of the third component produces an improvement in the resulting fit statistic significant at $\rm \geq 99 \,\% $ confidence level. 
In the whole sample, we found that the NIR excess blackbody component is required in 5 out of the 16 sources ($\rm \sim 31 \%$).



\subsection{Bolometric and Monochromatic luminosities}
\label{sec:bolo}
From the final SED best-fit, we derived several physical quantities of both the host galaxy and the nuclear source.  
The intrinsic quasar bolometric luminosity has been computed for each source by integrating the accretion disk emission in the range $ \rm 60 \, \AA-1 \, \mu m$ (despite Fig. \ref{fig:sed1} and \ref{fig:sed2} only showing wavelengths above $\rm \sim 500 \, \AA$).\\
In particular:

\begin{equation}
L_{bol}=4 \pi d_L^2 \int_{60 \, \AA}^{1 \, \mu m} A\, f_{AD} \; \; [erg/s]
\end{equation}

\noindent
where $d_L$ is the luminosity distance and $f_{AD}$ is the accretion disk emission component with $ A $ relative normalization.  In the computation of the bolometric luminosity, we do not account for the soft X-ray emission, which however has been demonstrated to be negligible. In fact, the bolometric corrections, i.e. the ratio between the bolometric luminosity and the luminosity at a specific wavelength, found in the soft X-ray band are quite large, of the order of 20-30, as in e.g. \citet{Marconi2004} and \citet{Lusso2012}.\\
Similarly, the quasar monochromatic luminosities at $2500 \, \AA$, $4500 \, \AA$ and $\rm 6\, \mu m$ and the one at $\rm 158\, \mu m$ (which corresponds to CII emission line), have been computed as:
\begin{equation}
\lambda L_{\lambda}=4 \pi d_L^2 \; \lambda \, f(\lambda)  \; \; [erg/s]
\end{equation}
where $\rm f(\lambda)$ has been obtained by interpolating the best fit model accretion disk plus torus template (for $\rm L_{2500 \, \AA}$, $\rm L_{4500 \, \AA}$ and $\rm L_{6\, \mu m}$) and the total template (for $\rm L_{158\, \mu m} $), and properly correcting for dust absorption, considering the value of the corresponding intrinisic template.

The quasar bolometric  and  the monochromatic luminosities computed as described above are reported in Table \ref{tab:lumi}. As expected, these quasars show very high bolometric luminosities (in the range $\rm 10^{47} - 10^{48} erg/s$), populating the brightest end of the luminosity function. \\

\begin{figure}
\includegraphics[scale=0.4]{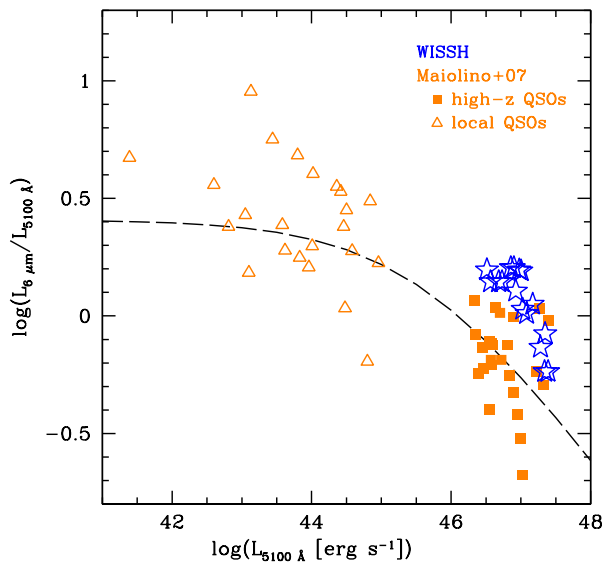}
\caption{MIR-to-optical continuum luminosity versus optical continuum luminosity. The WISSH quasars are the blue stars in the plot, while in orange the sample from \citet{Maiolino2007} of high-z (squares) and local (triangles) sources. The dashed black line is the relation derived by \citet{Maiolino2007}.}
\label{fig:cf}
\end{figure}

It is well known that the optical continuum luminosity can be used as a proxy of the quasar luminosity while the MIR flux is directly linked to the circumnuclear hot dust emission. Moreover, the dust covering factor for type I AGN can be obtained from the ratio between the thermal MIR emission and the primary quasar radiation. Under these assumptions, a plot as that in Figure \ref{fig:cf} has been showed by \citet{Maiolino2007} to recognize a mild trend of the covering factor with the quasar luminosity, i.e. decreasing with optical luminosity. A strong trend with luminosity emerges also from our sources (blue stars in Figure \ref{fig:cf}), which however lie slightly above (of $\rm \sim 0.2$ dex) the relation derived by \citet{Maiolino2007}. This is not surprising since the WISSH quasars have been selected to be the most luminous in the $\rm 22 \, \mu m$ band which corresponds to about $\rm \sim 6 \, \mu m $ rest-frame given their redshift.

\begin{table*}
\begin{center} 
\begin{tabular}{cccccc}
\toprule\toprule
$\rm \tiny{N}$ & $ \rm \tiny{log(L_{bol})}$ & $ \rm \tiny{log(L_{2500\AA})}$ & $ \rm \tiny{log(L_{4500\AA})}$ & $\rm \tiny{log(L_{6\, \mu m})}$ & $\rm \tiny{log(L_{158\, \mu m})}$\\
 &  [$\rm erg/s$] &  [$\rm erg/s$] &  [$\rm erg/s$] & [$\rm erg/s$] & [$\rm erg/s$] \\
\midrule\midrule
  \phantom{1}1 &  \tiny{47.54}  &  \tiny{47.05}  &   \tiny{46.93}  & \tiny{47.10}	& \tiny{45.73}	\\
  \phantom{1}2 &  \tiny{47.64}  &  \tiny{47.15}  &   \tiny{47.03}  & \tiny{47.19}	& \tiny{45.84}    \\
  \phantom{1}3 &  \tiny{47.65}  &  \tiny{47.16}  &   \tiny{47.04}  & \tiny{47.05}	& \tiny{45.77}    \\
  \phantom{1}4 &  \tiny{48.03}  &  \tiny{47.54}  &   \tiny{47.41}  & \tiny{47.15}	& \tiny{46.05}    \\
  \phantom{1}5 &  \tiny{47.80}  &  \tiny{47.31}  &   \tiny{47.18}  & \tiny{47.22}	& \tiny{46.05}    \\
  \phantom{1}6 &  \tiny{47.36}  &  \tiny{46.87}  &   \tiny{46.75}  & \tiny{46.87}	& \tiny{46.00 }   \\
  \phantom{1}7 &  \tiny{47.98}  &  \tiny{47.49}  &   \tiny{47.36}  & \tiny{47.10}	& \tiny{45.52}    \\
  \phantom{1}8 &  \tiny{47.31}  &  \tiny{46.82}  &   \tiny{46.70}  & \tiny{46.83}	& \tiny{46.15}   \\
  \phantom{1}9 &  \tiny{47.20}  &  \tiny{46.71}  &   \tiny{46.58}  & \tiny{46.71}	& \tiny{45.65}    \\
   10 &  \tiny{47.62}  &  \tiny{47.13}  &   \tiny{47.00}  & \tiny{47.17}	& \tiny{45.94}    \\
   11 &  \tiny{47.15}  &  \tiny{46.66}  &   \tiny{46.53}  & \tiny{46.71}	& \tiny{46.15}    \\
   12 &  \tiny{47.92}  &  \tiny{47.43}  &   \tiny{47.30}  & \tiny{47.15}	& \tiny{45.45}    \\
   13 &  \tiny{47.56}  &  \tiny{47.07}  &   \tiny{46.94}  & \tiny{47.03}	& \tiny{45.42}    \\
   14 &  \tiny{47.98}  &  \tiny{47.49}  &   \tiny{47.36}  & \tiny{47.27}	& \tiny{45.83}    \\
   15 &  \tiny{47.71}  &  \tiny{47.22}  &   \tiny{47.10}  & \tiny{47.09}	& \tiny{45.54}    \\
   16 &  \tiny{47.50}  &  \tiny{47.01}  &   \tiny{46.88}  & \tiny{47.06}	& \tiny{45.69}    \\
\bottomrule
\end{tabular}
\caption{Quasar Bolometric luminosity and Monochromatic luminosities at $\rm 2500 \, \AA$, $\rm 4500 \, \AA$, $\rm 6 \, \mu m$ and $\rm 158 \, \mu m$ }
\label{tab:lumi} 
\end{center}
\end{table*} 

\subsection{Infrared Luminosities and Star formation rates}
An interesting parameter to derive is the IR luminosity of the host galaxy component, which is due to the reprocessed UV stellar emission by dust and indeed depends on both the physical properties of the dust (i.e. mass) and of the incident radiation (i.e. star formation and the quasar, see discussion in Section \ref{sec:bianchi}). 

We have computed the star formation rate (SFR)  using the  relation by \citet{Kennicutt1998} scaled to a Chabrier IMF:
\begin{equation}
SFR(M_\odot / yr) = 10^{-10} L_{IR \, 8-1000 \, \mu m}^{Host}\; (L_\odot)
\end{equation}
where the IR host galaxy luminosity ($\rm L_{IR}^{Host}$) has been obtained by integrating the best fit modified blackbody in the range \, $\rm 8-1000 \, \mu m$:
\begin{equation}
L_{IR \, 8-1000 \, \mu m}^{Host}=4 \pi d_L^2 \int_{8\, \mu m}^{1000\,\mu m} B\, f_{CD} \; \; [erg/s]
\end{equation}
Both $\rm L_{IR \, 8-1000 \, \mu m}^{Host}$ and SFR are given in Table \ref{tab:infraprop}, while $B$ is the normalization related to the CD template. The values of SFR found for such sources are extremely high ($\rm >400 \, M_\odot / yr$ and up to $ \rm \sim 4500 \, M_\odot / yr$). \\


\begin{figure}
\includegraphics[scale=0.4]{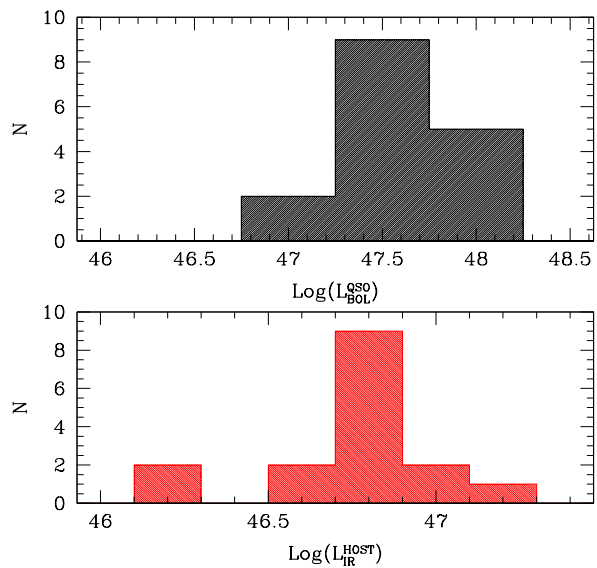}
\caption{Distribution of the bolometric luminosities (upper panel) and of the host IR luminosities (lower panel) of the sample.}
\label{fig:histolumi}
\end{figure}

\begin{table} 
\begin{center}
\begin{tabular}{cccc}
\toprule\toprule
$\rm \tiny{N}$ & $\rm \tiny{log (L_{IR\,8-1000 \, \mu m}^{Host}) } $ & $\rm \tiny{SFR} $ & $\rm \tiny{log (L_{FIR\,40-120 \, \mu m}^{Host}) } $ \\
 &  [$\rm erg/s$] &  [$ \rm M_\odot / yr $] &  [$\rm erg/s $] \\
\midrule\midrule
\phantom{1}1  &  \tiny{< 46.72} &  \tiny{< 1400} &  \tiny{< 46.59}  \\
\phantom{1}2  &  \tiny{\phantom{1} 46.87} &  \tiny{\phantom{2} 2000} &  \tiny{\phantom{1} 46.73}  \\	
\phantom{1}3  &  \tiny{\phantom{1} 46.55} &  \tiny{\phantom{3} 930} &  \tiny{\phantom{1} 46.44}  \\
\phantom{1}4  &  \tiny{\phantom{1} 46.83} &  \tiny{\phantom{2} 1800}   &  \tiny{\phantom{1} 46.72}   \\
\phantom{1}5  &  \tiny{\phantom{1} 47.01} &  \tiny{\phantom{2} 2700}   &  \tiny{\phantom{1} 46.88}   \\
\phantom{1}6  &  \tiny{\phantom{1} 46.90} &  \tiny{\phantom{2} 2100}   &  \tiny{\phantom{1} 46.78}   \\
\phantom{1}7  &  \tiny{\phantom{1} 46.54} &  \tiny{\phantom{3} 910} 	   &  \tiny{\phantom{1} 46.41}  \\
\phantom{1}8  &  \tiny{\phantom{1} 47.24} &  \tiny{\phantom{2} 4500}   &  \tiny{\phantom{1} 47.09}   \\
\phantom{1}9  &  \tiny{\phantom{1} 46.70} &  \tiny{\phantom{2} 1300}   &  \tiny{\phantom{1} 46.56}  \\
10  &  \tiny{\phantom{1}46.77} &  \tiny{\phantom{2} 1600}   &  \tiny{\phantom{1} 46.66}   \\
11  &  \tiny{\phantom{1}46.92} &  \tiny{\phantom{2} 2200}   &  \tiny{\phantom{1} 46.81}   \\
12  &  \tiny{< 46.21} & \tiny{< \phantom{1}410}   &  \tiny{< 46.10}   \\
13  &  \tiny{\phantom{1}46.90} &  \tiny{\phantom{2} 2000}    &  \tiny{\phantom{1} 46.61}  \\
14  &  \tiny{\phantom{1}46.85} &  \tiny{\phantom{2} 1800}   &  \tiny{\phantom{1} 46.71}    \\
15  &  \tiny{\phantom{1}46.27} &  \tiny{\phantom{3} 490}    &  \tiny{\phantom{1} 46.17}   \\
16  &  \tiny{\phantom{1}46.79} &  \tiny{\phantom{2} 1600}   &  \tiny{\phantom{1} 46.64}    \\
 
\bottomrule
\end{tabular}
\caption{Infrared luminosity computed by integrating the host emission from $\rm 8 \, \mu m$ to $\rm 1000 \, \mu m $, related star formation rates as from \citet{Kennicutt1998} and FIR luminosity of the host component from $\rm 40 \, \mu m$ to $\rm 120 \, \mu m $.}
\label{tab:infraprop} 
\end{center}
\end{table}

These derived values of SFR assume that all the FIR emission comes from young stars. However, we know that this is not completely true since the quasar emission might contribute to it \citep{Schneider2015,Symeonidis2017}. For this reason, the values of SFR reported in the table have to be considered as upper limits. A more detailed discussion on this can be found in Section \ref{sec:bianchi}, where the quasar contribution to the FIR radiation has been accounted for in a statistical way.   

\subsubsection{Quasar contribution to the FIR emission}
\label{sec:bianchi}
While there is broad consensus that the strong NIR emission observed in quasars is due to the central source, the debate is still on whether the FIR emission is mainly powered by the starburst, or whether quasars are required as an additional energy source. \\
We estimated the possible contribution of the quasar to the heating of the dust in the host galaxy 
using the same approach as in \citet{Schneider2015}. For the z=6.4 quasar SDSS J1148+5251, they
modelled the dust in the host with a spheroidal inhomogeneous distribution extending from 
just outside the quasar dusty torus up to a few kpc. A central source was included, with a 
bolometric luminosity sufficient to match the observed optical data; the quasar SED was chosen
from a few templates including the contribution of the dusty torus in the MIR. 
The heating of the dust in the host by the central source (and by an additional contribution from
stars in the galaxy) was computed using the radiative transfer code TRADING (Bianchi 2008), assuming
the typical properties of dust in the Milky Way (Draine \& Li 2007). From their simulations,
\citet{Schneider2015} concluded that dust in the host galaxy, heated by the IR radiation from the
dusty torus, can contribute significantly (at least for $30 \%$ and up to $70 \%$) to the FIR SED of the quasar-host system.

We repeated the same procedure here, for two extreme sources in our sample: the least 
(SDSS J123714.60+064759.5) and the most (SDSS J074521.78+473436.1) luminous ones. Given the importance of such an estimate, we will apply the same procedure for the whole sample.
For the two sources in this work, we produced TRADING radiative transfer simulations, choosing as the central source the
nuclear and torus emission best-fit templates obtained from the optical/NIR data as described in Section \ref{sec:3}. We adopted the same dust distribution and mass as in Schneider et al. (2015): as we will see in the next Section, the dust
masses in our sample are of the same order of that for SDSS J1148+5251, studied by \citet{Schneider2015}. By comparing the modelled 
and observed SED, we found that the quasar contribution to the FIR fluxes is about 43$\%$ in the least luminous source raising to $60\%$ for the most luminous one. This points towards a mild trend with luminosity. However, considering that the bolometric luminosities are homogeneously distributed, as visible in Figure \ref{fig:histolumi}, we can assume an average quasar contribution of $\sim$50\% to the total FIR luminosity, which also accounts for the uncertainties in the radiative transfer model. The quasar-corrected infrared luminosities and SFR can be therefore derived simply dividing the values given in Table \ref{tab:infraprop} by a factor of 2.

\subsection{Star formation rate vs Black Hole accretion}
\label{sec:sfr}

\begin{figure*}
\begin{center}
\includegraphics[scale=0.999]{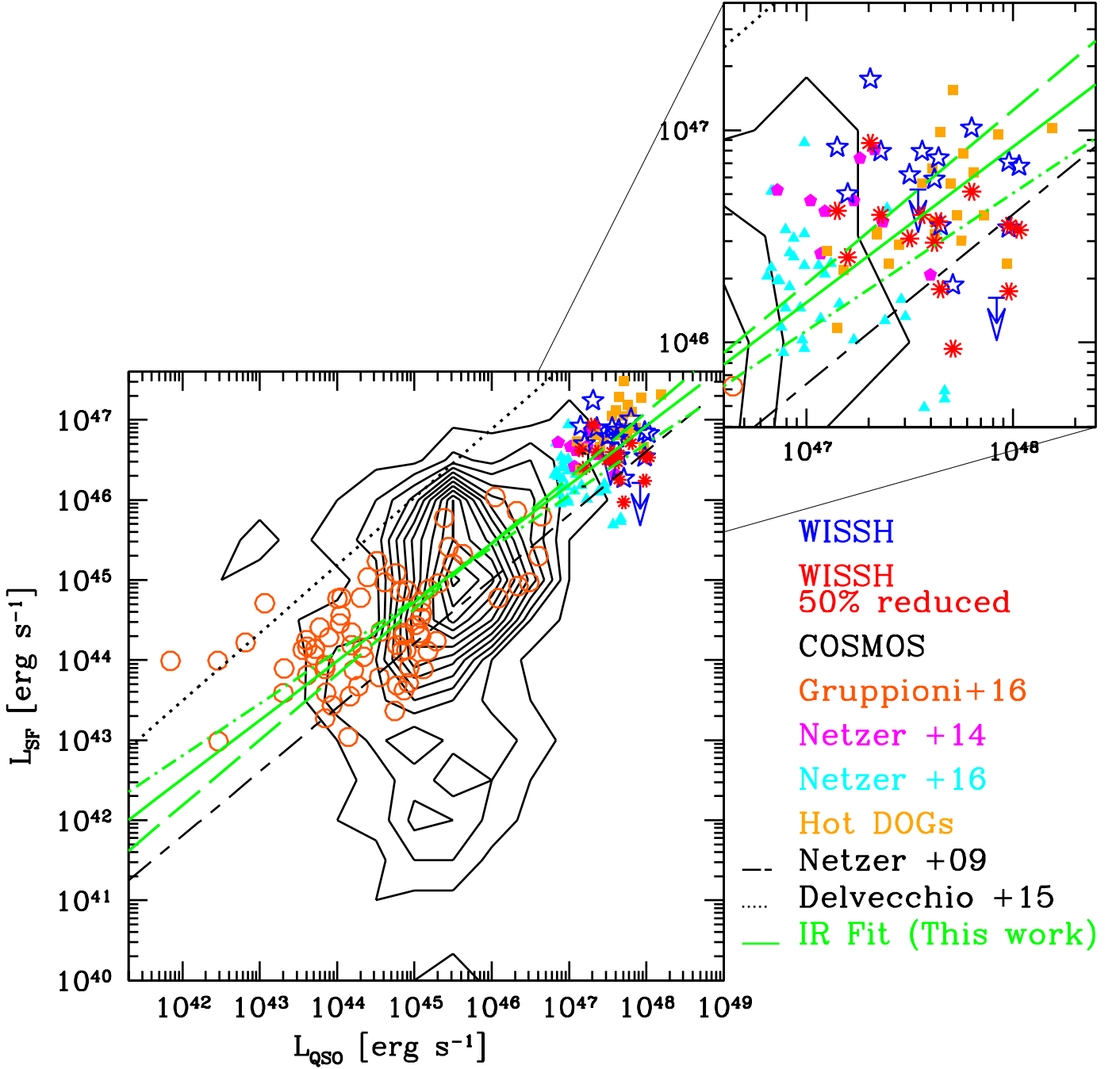}
\caption{Star formation luminosity vs quasar bolometric luminosity for the WISSH quasars in comparison with other samples. The 16 WISSH quasars are in blue stars in the plot, while in red asterisks are shown the values of the SF luminosity reduced by a factor of about 2 to account for the quasar contribution to the FIR. The same correction has been applied to the samples from \citet{Netzer2014} and \citet{Netzer2016} and \citet{Lulu2016}, which are of the same luminosity range.}
\label{fig:lsf}
\end{center}
\end{figure*}

Figure \ref{fig:lsf} shows the IR luminosity due to star formation (hereafter SF luminosity) versus the quasar bolometric luminosity. As expected,  the WISSH quasars (blue stars) populate an extreme region of this plane. In the same figure we also show as red asterisks, the values of SF luminosity statistically corrected for the quasar contribution. 
Such extreme values are in perfect agreement with the results found for quasars at high z with similar properties, as the two samples presented by \citet{Netzer2014} and \citet{Netzer2016}, and the Hot Dust Obscured Galaxies \citep[from the work by][]{Lulu2016}, shown in Fig. \ref{fig:lsf} as magenta pentagons, cyan triangles and orange squares, respectively. For three objects in our sample, in common with \citet{Netzer2016}, we can directly  compare the values of SFR, i.e., SDSS J020950.71-000506.4 (2)
with a SFR of $ \rm 1700 \, M_{\odot}/yr $ from \citet{Netzer2016} and of $ \rm 2000 \, M_{\odot}/yr $ from our analysis, SDSS J123714.60+064759.5 (11) having $ \rm 2700 \, M_{\odot}/yr $ and $ \rm 2200 \, M_{\odot}/yr $ in the two works, and SDSS J212329.46-005052.9 (15) with $ \rm 300 \, M_{\odot}/yr $ and $ \rm 490 \, M_{\odot}/yr $, respectively.\\
Previous studies about the presence of any possible correlation between the star formation activity and the quasar luminosity led to discrepant results. Several works based on X-ray data revealed the presence of a robust correlation at high redshift and quasar luminosity, which weakens or disappears for sources at lower redshift and luminosity. In particular \citet{Stanley2015}, \citet{Harrison2012} and \citet{Rosario2012}, dealing with data obtained via stacked photometry, found a flat distribution of $\rm L_{SF} - L_{QSO}$. However, \citet{Dai2015}, for a sample of FIR-detected quasars at $\rm 0.2 < z < 2.5$, found a significant trend between these two parameters over four orders of magnitude in luminosity. As noted in several previous works \citep[see e.g.][]{Netzer2016}, these discrepancies might be attributed to different selection criteria and/or methods of analysis. In particular, \citet{Dai2015} have been able to reconcile their results with those obtained by \citet{Stanley2015}, by considering stacked IR values. \\
Given the limited luminosity range spanned by our sample, we are not able to unveil any possible $\rm L_{SF} - L_{AGN}$ trend. In Fig. \ref{fig:lsf} we report the local sample of IR-selected galaxies from \citet{Gruppioni2016} shown as brown circles and the quasar sample from the COSMOS survey \citep{Bongiorno2012}, reported with density contours. The combination of such samples allows us to span a wide range of luminosity and redshift. Very high $\rm L_{QSO}$ objects are indeed missing from studies based on small fields, like COSMOS, which do not properly sample the bright end of the quasar luminosity function. What is clearly visible from the plot, is that at high luminosities (both bolometric and SF luminosity), there is a smaller dispersion in luminosity than what found for less luminous samples.  Quantitatively, the high-luminous sources have a dispersion which is 2.15 times smaller than the low-luminous ones. However, it is worth noting that all the aforementioned samples are \textit{Herschel}-detected, except for COSMOS. \\

\begin{figure*}
\begin{center}
\includegraphics[scale=0.7]{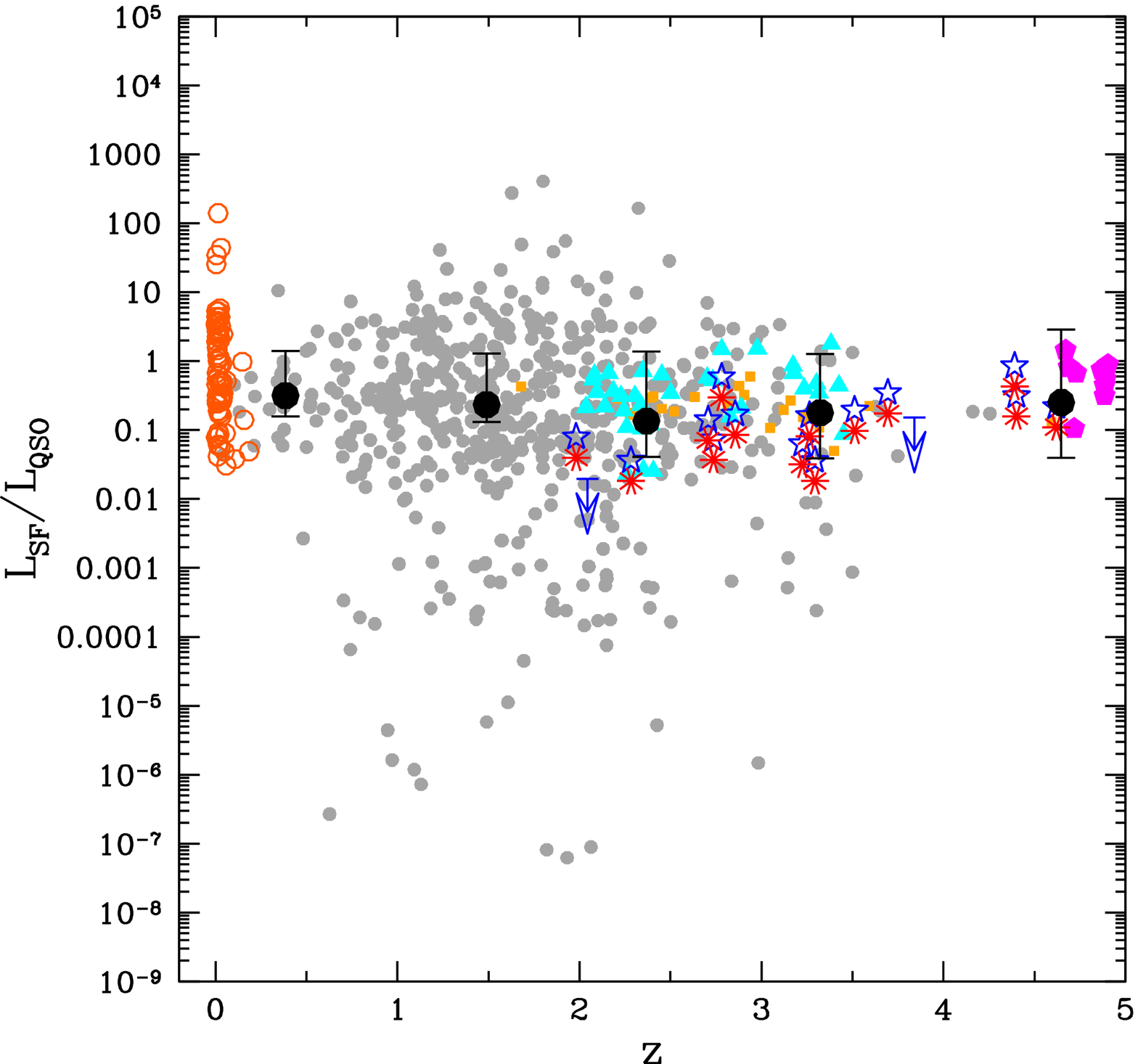}
\caption{Ratio between SF and quasar bolometric luminosities as a function of redshift for the same sources presented in Fig. \ref{fig:lsf}. Black points represent the median value in a redshift bin.}
\label{fig:lsfz}
\end{center}
\end{figure*}

We have investigated whether there is an hidden redshift dependence in the relation $\rm L_{SF} - L_{QSO}$. Fig.\ref{fig:lsfz} shows the ratio between the SF and the quasar bolometric luminosity versus z, for the same samples of  Fig. \ref{fig:lsf} (with COSMOS in grey points). Black points represent the median values in bins of redshift with associated errors, which have been computed considering the differences between the first and third quartiles and the median value (or second quartile) of the set of data. As is clearly visible, there is no sign of strong trend with redshift, which allows us to use samples at different redshift to derive a relation between the SF and the quasar activity for the IR sources. \\
Here we concentrate on the \textit{Herschel}-detected samples, i.e.  the local sample from \citet[][orange squares]{Gruppioni2016}, the $\rm z>2$ samples from \citet[][magenta pentagons]{Netzer2014} and \citet[][cyan triangols]{Netzer2016}, the Hot DOGs from \citet[][orange squares]{Lulu2016} and the WISSH ones (red asterisks) at $\rm 1.9<z<4.6$. For these samples we have obtained a fit which is reported in green solid line in Fig. \ref{fig:lsf}, and for which we have assumed the same correction for the quasar contamination to the FIR (50\%) also for the measurements belonging to the samples from \citet{Netzer2014} and \citet{Netzer2016} and from \citet{Lulu2016}, which have similar properties than ours. On the contrary, we cannot make the same assumption for the objects by \citet{Gruppioni2016}: these are local sources at lower luminosity, therefore the quasar contribution to the FIR may be different compared to high-luminous sources. To be more accurate, we calculated the bisector of the two regression lines first treating $\rm L_{QSO}$ (dashed green line) as the independent variable and then considering  the star-formation luminosity $\rm L_{SF}$ as the independent variable (dot-dashed green line). The bisector, the green straight line in the plot, has a dispersion of $\rm 0.39 $ dex and presents the following form:\\

\begin{equation}
\rm log \left( \frac{L_{SF}}{10^{44} erg/s} \right) = log \left( \frac{L_{QSO}}{1.82 \times 10^{44}erg\,s^{-1}}\right)^{0.73}
\end{equation}
\noindent
and can be directly compared with the relation found by \citet{Delvecchio2015} ($\rm L_{SF} \propto L_{BOL}^{0.85}$) for their SF-dominated sources (i.e. those with $\rm L_{SF}>L_{QSO}$) at $\rm 1.5<z<2.3$ and with that obtained by \citet{Netzer2009} for quasar-dominated sources ($\rm L_{SF} \propto L_{BOL}^{0.78}$). \\

As expected, the relation found differs significantly if compared to the one from \citet{Delvecchio2015}, while it is in much better agreement with the one derived by \citet{Netzer2009}, especially at the high luminosity end. 
At lower luminosities, on the contrary, the newly derived relation is flatter compared to the \citet{Netzer2009} one, with a slope of 0.73 instead of 0.78. The difference can be attributed to the fact that our relation has been derived considering only \textit{Herschel}-detected sources, while the relation published by \citet{Netzer2009} is mainly based on  optically selected  SDSS sources.\\

\subsection{Dust Temperatures and Masses}
\label{sec:masses}

\noindent
\textbf{(i) Cold Dust}\\ 

Under the assumption that dust emissivity can be described as a simple frequency power law, and that dust is in thermal equilibrium, from the result of the SED fitting we can also derive some dust properties, in particular its mass and temperature. More in detail, the FIR emission is linked to the cold dust component present on galaxy scale while the NIR emission is related to the hot dust component close to the nucleus in the torus. 
Following Wien's displacement law, we know that the peak of the blackbody component is strictly connected to the temperature of the emitting dust grains. 
We have already underlined the fact that the real SEDs of galaxies could not be described by a single-temperature modified blackbody because grains of different sizes and shapes give birth to a non-trivial distribution of temperatures. However, for simplicity, a blackbody (or a modified blackbody) component or a grid of templates (as \citealt{Chary2001} and \citealt{Dale2014}) also characterized by a single-temperature emission, is commonly used to describe the IR radiation \citep[see e.g.][]{Leipski2013,Netzer2016}.\\

\begin{figure*}
\begin{center}
\includegraphics[scale=0.75]{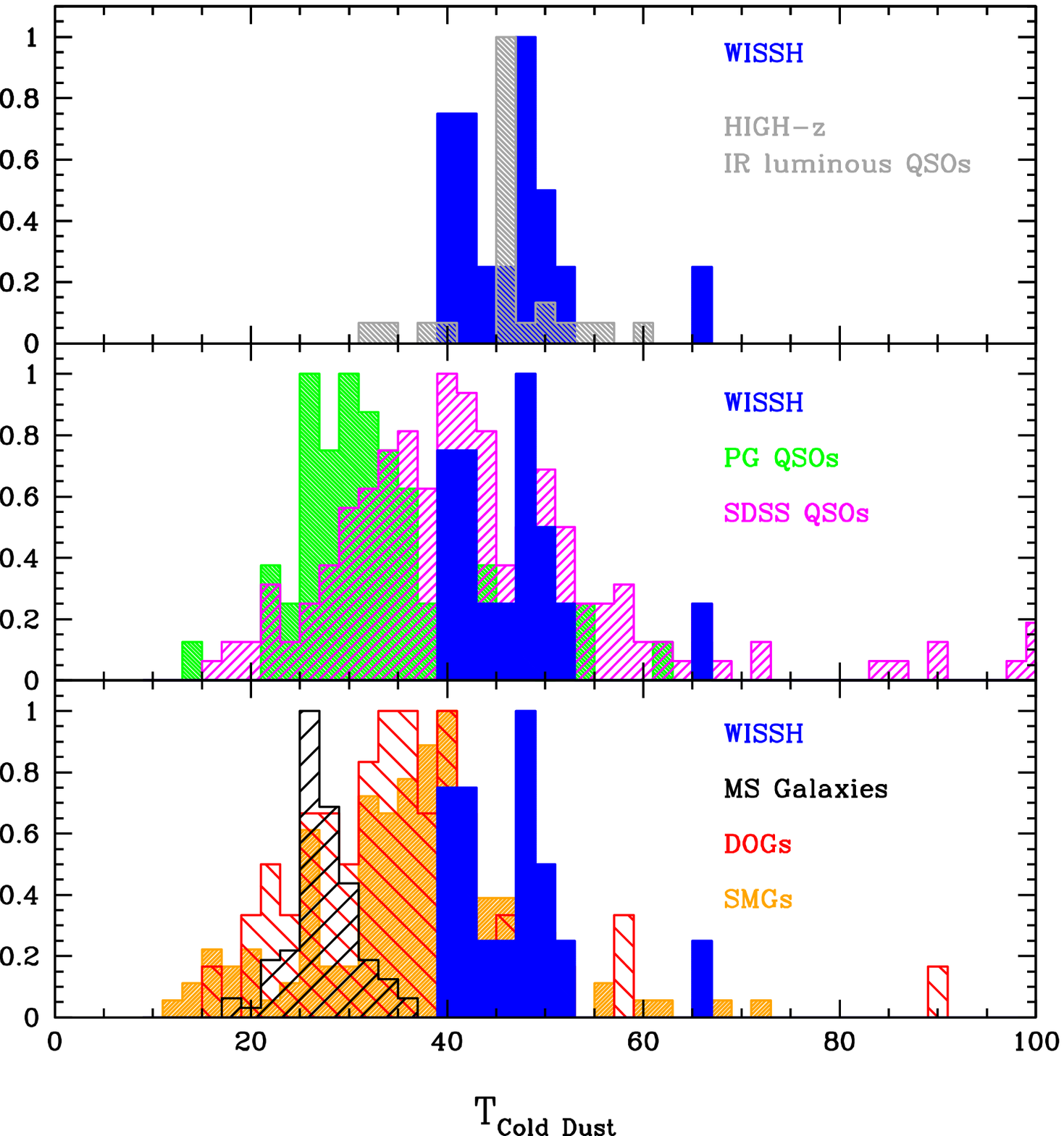}
\caption{Temperature distribution of the cold dust for the WISSH quasars (blue) compared to: high-z IR luminous quasars from \citet{Beelen2006,Leipski2014,Valiante2011} (upper panel);  local PG quasars from \citet{Petric2015} and SDSS quasars at z$<$4.7 from \citet{Ma2015} (central panel); and galaxies from \citet{Magdis2012, Santini2014,Melbourne2012,Chapman2004,Magnelli2014,Miettinen2015} (lower panel). Note that all these works have assumed different values of both the coefficient of the modified blackbody $\rm \beta$ and the dust absorption coefficient $\rm k_0$.}
\label{fig:histo_cold}
\end{center}
\end{figure*}

Figure \ref{fig:histo_cold} shows the distribution (normalized for the maximum value) of the temperatures of the cold dust component found for our objects (blue histogram), ranging from about $\rm T=40 \,K$ to $\rm T=50 \,K$, in good agreement with the results of previous studies of high-redshift, IR luminous quasars \citep[e.g.][]{Beelen2006,Leipski2014,Valiante2011}, as visible in the upper panel. Central and lower panels show, as comparison, the distribution of cold dust temperatures found for different classes of objects. It is important to note that all these works have assumed different values of both the coefficient of the modified blackbody $\rm \beta$ and the dust absorption coefficient $\rm k_0$. 
In the central panel are reported the WISSH sources compared to the local ($\rm z<0.5$) PG quasars from \citet{Petric2015} and the SDSS optically selected quasars at $\rm z<4.7$ from \citet{Ma2015}. The different IR luminosity range directly translates into a different dust temperature distribution,  i.e.  while the less luminous PG quasars have lower temperatures (in the range 20 K - 50 K) compared to the WISSH sample, the SDSS objects show a wider distribution, with the bulk peaking at higher value ($\rm \sim 40 \,K$) than the PG quasars, and with the warmer tail of the distribution comparable to ours. 
For the same reason, WISSH quasars show dust temperature much higher compared to the bulk of the galaxy population, as clearly seen in the lower panel of Fig. \ref{fig:histo_cold}, where they are compared to normal galaxies, i.e. the sample of main sequence galaxies from \citet{Magdis2012} and \citet{Santini2014}), the dust obscured galaxies (DOGs) from \citet{Melbourne2012} and the submillimiter galaxies (SMGs) from \citet{Chapman2004}, \citet{Magnelli2014} and \citet{Miettinen2015}. \\

Using the previously derived  $\rm L_{IR 8-1000 \, \mu m}^{Host}$, it is possible to obtain also a measure of the mass associate with the cold dust component (Cold M$_d$) using the formula derived by \citet{Hughes1997} \citep[see also][for a detailed derivation]{Berta2016}:

\begin{equation}
M_d = \frac{L_{IR \, 8-1000 \, \mu m}^{Host}}{4\pi \int{k_d(\nu)\,B(\nu,T_d)\,d\nu}}
\label{eq:dust}
\end{equation}

\noindent
valid under the assumption that the emission is optically thin with optical depth $\tau << 1$.

In the above formula, $\rm B(\nu,T_d)$ is the Planck function for dust with temperature $\rm T_d$, determined by the best fit; $\rm k_d(\nu)=k_0\,(\nu/\nu_0)^\beta$ with $k_0$ the absorption coefficient and $\nu_0$ the frequency at which $\tau$=1. Following \citet{Beelen2006}, here we assume $\rm k_0 = 0.4 \,\, [cm^2/g]$ and $\rm \nu_0 = 1200 \, \mu m$. The dust mass therefore depends on the IR luminosity (SFR) of the source and the temperature of the dust grains, i.e. for a given temperature, there is a linear relation between dust mass and SFR. 


The derived masses, reported in Table \ref{tab:mdust}, are of the order of $\rm 10^8$ M$_{\odot}$, in agreement with previous studies on high-z, IR luminous sources \citep[e.g.][] {Beelen2006,Wang2008,Banerji2017,Valiante2011,Ma2015}. \\

\noindent
\textbf{(ii) Hot Dust}\\

As already mentioned in Sec. \ref{sec:nirexcess}, some of our sources (31\%) show a NIR excess in their SED which requires an additional component to the classical torus models \citep[]{Stalevski2012, Stalevski2016}. The presence of such component has already been noted in several previous works on luminous quasars but its physical origin of this component is not yet clear. \citet{Edelson1986} were the first to note the need of an additional NIR component peaked at about $\rm 5 \, \mu m$, corresponding to a dust temperature of about $\rm 600 \,K$. More recently, \citet{Mor2009}, analyzing a sample of Spitzer-selected low-z quasars with luminosities ranging from $\rm 44.6<log(L_{bol})<46.8$, found that their objects require two additional dust components to fit the NIR emission. The first one (very hot dust component) with temperatures ranging from $\rm 900 \,K $ to $\rm 1800 \,K$, and a second one (warm dust component) at $\rm 200 \,K$. Similarly, \citet{Caballero2016} found a peak in the distribution of the very hot dust temperatures at about $\rm 1180 \,K$, and of the warm one at $\rm 400 \,K$, for a sample of luminous quasars at $\rm z<3$. 
While the warm component has been interpreted as dusty clouds in the Narrow Line Region, the very hot dust emission might be related to graphite clouds located at the edge of the dusty torus thus inside the sublimation radius of silicate (whose sublimation temperature is lower than for pure-graphite grains) but outside the Broad Line Region.  However, as the NIR SED of the total quasar emission (disk+dust) is strictly connected to the spectral shape assumed for the NIR spectrum of the disk, the resulting excess components are someway model dependent.\\

As  shown in Table \ref{tab:mdust}, for the WISSH sample, we found a NIR excess corresponding to dust with temperatures ranging from  $\rm T=650 \,K$ to $\rm T=850 \,K$. Such values are in agreement with the result from \citet{Edelson1986} but in between that found by \citet{Mor2009} and \citet{Caballero2016}.\\
It is worth noticing that this further component peaks at about $\rm 22 \, \mu m$ rest-frame, where the sources in the sample have been selected to be the brightest. This excess is a sign of the fact that our sources show more hot dust than that present in the torus models. Moreover, they may have a large covering factor, allowing clouds to intercept more continuum radiation.  



Furthermore, from the NIR excess emission component, we have also estimated the mass of the additional hot dust found. 
Differently from the cold dust component, described as a greybody,  the hot dust emission excess has been fitted with a single-temperature blackbody. Therefore it is in principle not possible to compute the hot dust mass using the previous relation. However, following \citet{Jiang2006}, we approximated the hot dust emission as a greybody with $\rm k_0=2.06\, \times 10^3 [cm^2/g]$. Since in a greybody the emissivity $\alpha$ is below $1$, these values have to be considered as lower limits and are in agreement with the values found in the aforementioned work. 

Finally, we have also computed the total hot dust mass for all sources (Hot $\rm M_d^{TOT}$), by fitting the photometric points from the UV to the NIR with a combination of pure accretion disk \citep{Feltre2012} and single temperature black body templates to describe the whole NIR emission.\\
The values of the total hot dust masses, ranging from tens to hundreds of solar masses, are given in Table ~\ref{tab:mdust} together with the masses associated to the excess for the 5 sources which require such component.

\begin{table*} 
\begin{center}
\begin{tabular}{cccccc}
\toprule\toprule
N & $\rm Cold \, T_d \; [K] $ &  $\rm Cold \, M_d \; [10^8 M_{\odot}] $  & $\rm  Hot \, T_{d}^{Ex}  \; [K] $  & $\rm  Hot \, M_{d}^{Ex} \; [M_{\odot}]$ & $\rm Hot \, M_{d}^{TOT} \; [\rm M_{\odot}]$  \\
\midrule\midrule
\phantom{1}1  &  \tiny{48} & \tiny{1.9}  & -  	&  - 			  &	\tiny{130}	\\						
\phantom{1}2  &  \tiny{49} & \tiny{2.5 } & -  	&  - 			  &	\tiny{190}	\\		
\phantom{1}3  &  \tiny{43} & \tiny{2.5 } & -  	&  -              &	\tiny{240}	\\
\phantom{1}4  &  \tiny{41} & \tiny{6.1}  & -  	&  -              &	\tiny{\phantom{1}72} 	\\
\phantom{1}5  &  \tiny{47} & \tiny{4.3}  & -  	&  -              &	\tiny{\phantom{1}90} 	\\
\phantom{1}6  &  \tiny{45} & \tiny{4.3}  & \tiny{790}	&  \tiny{\phantom{1}57}      &	\tiny{\phantom{1}85} 	\\
\phantom{1}7  &  \tiny{49} & \tiny{1.2}  & -  	&  -              &	\tiny{\phantom{1}84} 	\\
\phantom{1}8  &  \tiny{51} & \tiny{4.6}  & \tiny{742}	&  \tiny{106}     &	\tiny{210}	\\
\phantom{1}9  &  \tiny{50} & \tiny{1.5}  & \tiny{786}	&  \tiny{\phantom{1}51}      &	\tiny{\phantom{1}51} 	\\
10 &  \tiny{43} & \tiny{4.1}  & -  	&  -              &	\tiny{156}	\\
11 &  \tiny{41} & \tiny{7.5}  & \tiny{664}	&  \tiny{134}     &	\tiny{170}	\\
12 &  \tiny{42} & \tiny{1.4}  & -  	&  -              &	\tiny{237}	\\
13 &  \tiny{66} & \tiny{0.5}  & \tiny{814}	&  \tiny{\phantom{1}54}      &	\tiny{\phantom{1}69} 	\\
14 &  \tiny{49} & \tiny{2.3}  & -  	&  -              &	\tiny{169}	\\
15 &  \tiny{41} & \tiny{1.9}  & -  	&  -              &	\tiny{196}	\\
16 &  \tiny{52} & \tiny{1.5}  & -  	&  -              &	\tiny{145}	\\

\bottomrule
\end{tabular}
\caption{Cold and Hot dust masses computed as described in Section ~\ref{sec:masses}. In the table, we report both the total hot dust mass and the excess found for our sources alone. The eleven sources with no NIR BB additional component have no estimates of the hot dust mass excess.}
\label{tab:mdust} 
\end{center}
\end{table*}

\section{Conclusions}
\label{sec:5}
In this paper we studied a sample of hyper-luminous quasars, selected by cross-correlating data from the SDSS and the WISE All-Sky Survey, and focusing on 16 sources with \textit{Herschel}/SPIRE data coverage. \\

\noindent
Thanks to the multi-wavelength coverage, we were able to analyze in detail the source SEDs, fitting the observed fluxes with a multi-components model, based on a combination of quasar and host-galaxy emission emerging in the IR (where the reprocessed stellar light is re-emitted by the dust). The quasar emission has been described using a combination of power-laws for the primary source \citep{Feltre2012}, and the model by \citet{Stalevski2016} for the smooth and clumpy dusty torus, while the galaxy cold dust emission in the infrared bands has been modeled as a modified blackbody. 
However, we found that, in some cases, our sources show peculiar features in their SED which cannot be described by the standard modelization, i.e. a further component, in the form of a pure blackbody, is  necessary to reproduce their NIR emission. 
Through the SED multi-components fitting method we were able to derive robust measurements of both the quasar (i.e. bolometric and monochromatic luminosities) and the host galaxy properties (i.e. dust mass, SFRs).\\

Our main findings can be summarized as follows:
\begin{itemize}

\item The WISSH quasars populate the brigthest end of the luminosity function with very high bolometric luminosities, i.e. $\rm L_{BOL} > 10^{47} erg/s$  and are hosted in galaxies with extremely high SFRs, up to $\rm 4500 \, M_{\odot}/yr$. Since the quasar light might contaminate the FIR emission,  following \citet{Schneider2015}, we estimated its possible contribution to the heating of the dust in the host galaxy, taking as test cases the least and the most luminous sources of the sample. We found that the quasar contribution to the FIR fluxes is about 43$\%$ in the least luminous source raising to $60\%$ for the most luminous one. Considering that the bolometric luminosities are homogeneously distributed, we therefore assumed an average quasar contribution of $\sim$50\% to the total FIR luminosity.  Even accounting for such correction, the SFRs of the studied sources still remain of the order of thousands $\rm M_{\odot}/yr$;\\

\item By combining our sample with both high-z hyper-luminous quasars \citep{Netzer2014,Netzer2016} and local ones from \citet{Gruppioni2016} and the quasar sample from the COSMOS survey \citep{Bongiorno2012}, from the $\rm L_{SF}-L_{QSO}$ plane it emerges a narrower (2.15 times smaller) dispersion at high quasar and star formation luminosities than what found for lower values of both bolometric and star formation luminosities. Using only the \textit{Herschel}-detected samples we derived a log-linear relation between the SF and the quasar luminosities, i.e.  $\rm L_{SF}\propto L_{QSO}^{0.73}$, flatter than the one derived for optically selected type-II AGN by \citet{Netzer2009}.\\

\item While most of the WISSH quasars are well described by a standard combination of accretion disk+torus and cold dust emission, for $\sim$ 31\% of them, an additional hotter component is required to reproduce the observed fluxes. The peak of this emission falls roughly at about $\rm 22 \, \mu m$ rest-frame, where the sources in the sample have been selected to be the brightest, and has temperatures ranging from $\rm T=650 \,K$ to $ \rm T=850 \,K$. \\

\item The temperature of the cold dust component has a peak at about $\rm 50 \, K$, in agreement with previous studies of high-z IR luminous quasars \citep[]{Beelen2006,Leipski2014,Valiante2011} and much higher 
compared to main sequence galaxies. The thermal emission of the cold dust component is associated to a dust mass of the order of $\rm 10^8$ M$_{\odot}$, which is in good agreement with previous works on high-z quasars \citep[e.g.][]{Beelen2006, Wang2008, Banerji2017, Valiante2011, Ma2015}.\\

The WISSH quasars lie in the most extreme region of the plane $\rm L_{SF}-L_{QSO}$. It is worth noting that the same locus is occupied by other hyper-luminous ($\rm L_{BOL} > 10^{47} erg/s$) quasars collected according to the level of extinction. We can place the WISSH quasars in between the heavily dust enshrouded sources such as the WISE-selected hot Dust Obscured Galaxies by \citet{Lulu2016}, \citet{Piconcelli2015} and \citet{Wu2012}, the NIR-selected heavily reddened quasars by \citet{Banerji2015} and \citet{Glikman2015}, and finally the blue optically selected luminous quasars \citep[e.g.][]{Netzer2016}. However, one should keep in mind that, while both the bolometric luminosity and the extinction are instant observables, the infrared-derived star formation luminosity is the result of the integration of bursts of star formation occured over billions of years, assuming that the SF remains constant over the lifetime of the burst  \citep[]{Leitherer1995}. Hence, from the star formation luminosity it is difficult to gain information about the ongoing star formation in the host galaxy. On the other hand, the extinction of a single galaxy is strictly linked to its merger-induced evolutionary sequence. Through the WISSH quasars we are therefore witnessing a very peculiar phase in the quasar life, during which the dust in the host has not been yet fully cleared up. This suggests that the typical evolution timescales of the quasar feedback are much shorter than those of the host galaxy star formation activity.\\

\noindent
Signatures of ionized outflows are ubiquitously detected in WISSH quasars \citep[see e.g.][Vietri et al. in prep]{Bischetti2017}. Future observations with ALMA and NOEMA will be fundamental to quantify the gas reservoir in the WISSH quasars and in revealing the presence of possible massive molecular outflows, allowing to obtain a comprehensive view of the outflow phenomenology in these hyper-luminous quasars, and test the impact of the quasar output on the host gas ISM and  its star formation efficiency.\\
An alternative scenario suggests that the action of the quasar outflow and the host star formation could not be geometrically connected. In this sense, future spatially-resolved observations will be useful to perform an accurate investigation of the ongoing star formation (via spectral analysis of the narrow $\rm H \alpha$ emission line, as in \citet{Carniani2015} and \citet{CanoDiaz2012}). \\
These quasars are also unique laboratories to study the properties of the interstellar medium at high redshift, as the metals and dust content of their host galaxies can provide insights on both the galaxy evolution process (e.g. SFH) and the quasar feedback effects \citep[e.g.][]{Valiante2011,Valiante2014}.

\end{itemize}

\begin{acknowledgements}
We thank Dr. J. Mullaney for his prompt referee report and for his valuable comments that improved the quality of the manuscript.\\
We are grateful to H. Netzer and C. Lani for the useful discussion on the infrared templates and the values of SFR. We also thank I. Delvecchio for the helpful conversation on the SED-fitting methodology and P. Santini for the help in the analysis of the dust masses. \\
A. Bongiorno and E. Piconcelli acknowledge financial support from INAF under the contract PRIN-INAF-2012. Raffaella Schneider and Rosa Valiante acknowledge support from the European Research Council under the European Union's seventh framework programme (FP7/2007-2013) ERC Grant Agreement 306476. L. Zappacosta acknowledges financial support2D under ASI/INAF contract I/037/12/0.\\

\end{acknowledgements}


%
\bibliographystyle{aa} 
\bibliography{fede} 
%

\end{document}